\newcommand{\beq}{\begin{equation}}
\newcommand{\eeq}{\end{equation}}
\newcommand{\bea}{\begin{eqnarray}}
\newcommand{\eea}{\end{eqnarray}}
\newcommand{\bear}{\begin{eqnarray*}}
\newcommand{\eear}{\end{eqnarray*}}

\newcommand{\rf}[1]{(\ref{#1})}
\newcommand{\scp}{\scriptsize}

\documentclass[prb,twocolumn,showpacs,floatfix]{revtex4}

\usepackage{graphicx}

\begin{document}

\title
{Density-functional formulations for quantum chains}
\author{Francisco C. Alcaraz and Klaus Capelle}
\affiliation{Instituto de F\'{\i}sica de S\~ao Carlos,\\
Universidade de S\~ao Paulo, 
CP 369, 13560-970, S\~ao Carlos, SP, Brazil}, 
\date{\today}

\begin{abstract}
We show that a lattice formulation of density-functional theory
(DFT), guided by renormalization-group concepts, can be used to obtain
numerical predictions of energy gaps, spin-density profiles, critical 
exponents, sound velocities, surface energies and conformal anomalies of 
spatially inhomogeneous quantum spin chains. To this end we (i) cast the 
formalism of DFT in the notation of quantum-spin chains, to make the powerful 
methods and concepts developed in {\em ab initio} DFT available to workers 
in this field; (ii) explore to what extent simple local approximations in 
the spirit of the local-density approximation (LDA), can be used to predict 
critical exponents and conformal anomalies of quantum spin models; 
(iii) propose and explore various nonlocal approximations, depending
on the size of the system, or on its average density in addition to the 
local density. These nonlocal functionals turn out to be superior to LDA 
functionals.
\end{abstract}

\pacs{75.10.Pq, 71.15.Mb, 75.10.Jm, 05.50.+q}


\maketitle


\section{Introduction}

In this work we show that a lattice formulation of density-functional theory
(DFT), guided by renormalization-group concepts, can be used to obtain
numerical predictions of energy gaps, spin-density profiles, critical
exponents and conformal anomalies of spatially inhomogeneous quantum spin
chains.

Density functional theory emerged over the years as an efficient and
powerfull tool for electronic-structure theory in condensed matter physics,
material science and quantum chemistry. In spite of these successful
applications of DFT to {\em continuum} systems,\cite{kohnrmp,dftbook} DFT is 
not much used for interacting {\em discrete} systems, i.e., systems where the
operators are defined on a lattice. Previous applications of DFT for discrete
systems were to the one-dimensional Hubbard model describing electrons in 
solids\cite{ah1,ah2,ah3,ah4,ah5} or atoms in optical lattices,\cite{ah6,ah7,ah8}
and to the Heisenberg spin-$\frac{1}{2}$ model in d-dimensions,\cite{ahe1,ahe2,ahe3,ahe4} employing local approximations similar to the local-density 
approximation (LDA) of {\em ab initio} electronic structure theory. The 
difficulty in treating such systems resides in the construction of a good 
functional. In Ref.~\onlinecite{ah3}, an LDA was constructed for the Hubbard 
model by exploiting the exact Bethe ansatz solution for the ground state of 
the infinite, spatially homogeneous, system. In Ref.~\onlinecite{ahe1},
LDA-like functionals for the Heisenberg model were obtained from an approximate 
spin-wave theory for the infinite homogeneous system, improved by adding 
numerical corrections from density-matrix renormalization-group calculations.

The purpose of this paper is to introduce a possible DFT formulation
appropriate for general quantum chains, in particular near their quantum
critical points.
The construction of approximate density functionals exploits two seminal
concepts from the theory of critical phenomena: the renormalization group
and the conformal invariance of critical systems. From the renormalization
group, it is known that the critical systems can be classified in universality
classes. In each of these classes the low-energy physics is described by
distinct homogeneous systems (fixed points) with correlation functions defined
by appropriate critical exponents. Moreover, the underlying field theory
describing most of these universality classes is conformally invariant. 
This symmetry classifies these classes by the conformal anomaly $c$,
or the central charge of the conformal algebra, and the critical exponents
are given in terms of the highest weights representations of this 
algebra.\cite{BPZ}
These facts lead us to expect that a {\em single} density functional
could describe the whole {\em family} of critical models of a given
universality class of critical behavior. In principle, the construction of
such functionals can proceed by exploiting the exact results provided along
the years by the Bethe ansatz\cite{bethe} in its several formulations.

It is believed that for each universality class of critical behavior there
is at least one exactly integrable quantum chain. The exact solution of that
quantum chain underlies the construction of the functional for several models
belonging to the universality class.
In order to test this general idea we restrict ourselves to one of the most 
important universality classes of critical behavior, namely, the $c=1$
conformal invariant models, with continuously varying critical exponents.
Models in this universality class are the Gaussian model,\cite{gaussian}
the anisotropic antiferromagnetic Heisenberg model with half-odd integer
spin,\cite{ab2,alc-moreo} the ferromagnetic version of the exact integrable
Taktajan-Babujian models,\cite{alc-martins} the anisotropic biquadratic 
spin $1$ chain,\cite{biqua} the Ashkin-Teller quantum chain,\cite{ab2} etc. 

As a representative of this class of models we are going to consider the 
exactly integrable spin-$\frac{1}{2}$ anisotropic Heisenberg model or the 
XXZ quantum chain. This model is considered a paradigm of integrability, and 
its critical exponents are known exactly.

Due to the conformal invariance, the leading finite-size corrections of the
eigenenergies of the quantum chains are ruled by the conformal anomaly and
the critical exponents of the critical chains.\cite{anomaly,cardy} The exact
knowledge of the ground state energy in the bulk limit and in the finite
geometry will allow us to construct good LDA approximations for
the functionals. Some of these LDAs will be good enough to reproduce the 
exact values of the conformal anomaly and critical exponents. Such LDA
approximations, although  applied in this paper only to the XXZ chain,
reveal interesting general features, that certainly will be helpful for other
discrete quantum systems. Our analysis reveals the essential ingredients and
the degree of accuracy we should demand from a density functional to obtain
from it reliable results for the conformal anomaly and critical exponents.

Once such a functional is available, numerical analysis of complex
Hamiltonians is greatly simplified. As an example, in the case we are mainly
concerned with below, the XXZ chain, full exact diagonalization becomes
hard for more than $\approx 12$ sites, Lanczos methods allow one to reach
about three times as much, whereas density-functional techniques can still
be applied for systems with thousands of sites, even if translational 
invariance is broken.

The layout of this papers is as follows. In Sec.~\ref{secII} we give
an introduction to the general framework of DFT for spin chains 
(Sec.~\ref{secIIA}) and to local approximations that make this 
framework useful in practice (Sec.~\ref{secIIB}). In Sec.~\ref{secIII} we
then describe the representative spin chain we are mainly concerned with in
this work: the XXZ model. Section~\ref{secIV} is devoted to the construction
of local approximations for the XXZ chain, and an exploration of their
performance for systems with periodic (Sec.~\ref{secIVA}), twisted
(Sec.~\ref{secIVB}) and open (Sec.~\ref{secIVC}) boundary conditions.
In Sec.~\ref{secV} we then go beyond local approximations, and construct
nonlocal functionals (Sec.~\ref{secVA}), which are applied to
systems with open boundary conditions (Sec.~\ref{secVB}) and in the
presence of spatially varying magnetic fields (Sec.~\ref{secVC}).
Section~\ref{secVI} contains our conclusions.

\section{Density-functional approach to quantum spin chains}\label{secII}

\subsection{General aspects of DFT for quantum spin chains}\label{secIIA}

Density-functional theory is a formally exact way to cast the
many-body problem in terms of densities instead of wave functions. For
a large class of Hamiltonians --- characterized by a stable ground state
and containing a density-like intensive variable coupled to an external
field --- the fundamental Hohenberg-Kohn theorem shows that the density
variable determines the ground state wave function, which implies that
all ground state observables are functionals of the density.\cite{kohnrmp,dftbook}

This general theorem does not tell one how to obtain these functionals,
how to calculate the density, what the physical nature of the external
field is --- and not even how to select the density variable in the
first place. All of these questions must be answered before applying
DFT to a specific class of problems. Here we attempt to do this for a
large family of spin Hamiltonians, by combining concepts from DFT with
renormalization-group ideas.

In practice, the density variable is typically chosen to represent the
charge or spin distribution in a continuous space or on a lattice.
The ground state energy becomes a functional of this distribution, which
is minimized to provide the ground state density. Of course, the energy
functional is not known exactly for nontrivial many-body systems.
Successful approximation schemes are based on the local-density concept,
which takes as an input the ground state energy of a spatially homogeneous
system, in which the density distribution is uniform. The corresponding
energy density is then evaluated point by point at the actual densities
of the inhomogeneous system under study.

While this means that DFT in the local-density approximation (LDA) requires
as an input a good solution to the spatial homogeneous system (which in
itself may be hard to obtain), it also implies that once that solution is
avaliable, the LDA can be used to obtain approximations to the ground state
energy and density distributions of inhomogeneous systems. In {\em ab initio}
applications of DFT, the homogeneous system is the electron liquid, the
external potential is produced by the nuclei, and the resulting inhomogeneous
systems are atoms, molecules and solids.\cite{kohnrmp,dftbook}

For model Hamiltonians defined on lattices, the homogeneous system is one
in which all sites are equivalent, and inhomogeneous systems can have
boundaries, impurities, confining potentials, or other translation-symmetry
breaking terms. 
Since model Hamiltonians are typically much better controlled than the 
{\em ab initio} one, much more information is available for constructing
the LDA and for testing its predictions. This additional information provides
a two-pronged opportunity: (i) Learn about inhomogeneous model Hamiltonians,
which have too few symmetries to be integrable and require too large matrices
to be numerically diagonalizable, by means of LDA calculations. (ii) Learn
about DFT, the LDA, and its improvements, by studying them in the context
of model Hamiltonians. Here we explore both aspects for spin Hamiltonians,
and inquire, in particular, if and how LDA-type functionals can be used to
predict the critical exponents and conformal anomalies associated with the
quantum critical points of these Hamiltonians.

\subsection{Local density functionals for quantum spin chains}\label{secIIB}

The family of quantum chains for which we are going to seek a DFT formulation, 
describes quantum spin operators attached to $L$ sites ($i=1,\ldots,L$) of a 
one dimensional lattice, with Hamiltonian
\beq \label{2.2}
H = H_0 + V^{\mbox{\scp ext}}.
\eeq
The operator $H_0$ may contain, e.g., the kinetic energy operator and the 
static interaction energy among the spins on distinct sites, while 
$V^{\mbox{\scp ext}}$ is the on-site potential.

Here, we consider Hamiltonians of the form
\rf{2.2} that commute with the global charge $\hat{N}=\sum_{i=1}^L \hat{n}_i$,
whose eigenvalues take a discrete set of values ($N=0,1,2,\ldots$).
Moreover the on-site potential in \rf{2.2} is a function of the local density
operators $\hat{n}_i$, having the general form
\beq \label{2.3}
V^{\mbox{\scp ext}} = \sum_{i=1}^L v_i^{\mbox{\scp ext}}\hat{n}_i,
\eeq
where $v_i^{\mbox{\scp ext}}$ ($i=1,\ldots,L$) are arbitrary site-dependent
couplings that fix the profile of  the spacial inhomogeneities of the quantum
chain. Examples of models in this family are the spin $S$ Heisenberg models,
where $\hat{N}$ is related to the $S^z$-magnetization, and the 
Hubbard\cite{lieb-wu} and t-J models, where $\hat{N}$ is related 
to the number of fermions.

For this class of models we can separate the Hilbert space associated to
\rf{2.2}
into block disjoint sectors labelled by the eigenvalues $N=0,1,2,\ldots$.
The variational principle, restricted to a given sector of total charge $N= 
\sum_{i=1}^L n_i$, give us the lowest eigenenergy on the sector
\beq \label{2.4}
E_0 = \mbox{min}_{|\Psi>} <\Psi^{(N)}| H |\Psi^{(N)}> = <\Psi_0^{(N)}| H |
\Psi_0^{(N)}>,
\eeq
where $|\Psi^{(N)}_0>$ is the eigenfunction corresponding to this lowest
eigenenergy.

The minimization procedure of \rf{2.4} can be done conveniently in two steps
by following the constrained-search approach of Levy\cite{levy} and
Lieb.\cite{lieb} In the first 
step we consider, for a given density distribution $n_i$ ($i=1,\ldots,L$),
restricted to $\sum_{i=1}^Ln_i = N$, the ensemble of states 
$\{|\Psi>\}_{n_1,\ldots,n_L}$ such that $<\Psi|\hat{n}_i|\Psi> = n_i$
($i=1,\ldots,L$). On this ensemble we minimize $<\Psi|H|\Psi>$, i.e.,
\beq \label{2.5}
\mbox{min}_{|\Psi>_{n_1,\ldots,n_L}} <\Psi|H|\Psi> 
= F_I[n] + \sum_{i=1}^L v_i^{\mbox{\scp ext}}n_i,
\eeq
where the special form of \rf{2.3} was used, and
\beq \label{2.6}
F_I[n] =
\mbox{min}_{|\Psi>_{n_1,\ldots,n_L}} <\Psi| H_0|\Psi>
\eeq
is in general a functional of the density $n$. On a lattice, $F_I[n]$ reduces 
to a function of the $L$ independent variables $n_1,\ldots,n_L$, but for
consistency with the {\em ab initio} literature we continue to employ the 
expression 'functional'. For notational convenience, we below frequently 
employ $n$ in the argument of a functional, i.e., between square brackets,
to denote the entire discrete distribution of values $n_1,\ldots,n_L$. 
When used on its own, or as an argument of a simple function of one variable, 
$n$ represents just one value, which can be site dependent ($n_i$) or
constant ($n=N/L$). 

In the second step we minimize \rf{2.5} over all densities
satisfying
$\sum_{i=1}^L n_i =N$, i.e.,
\beq \label{2.7}
\delta \left[F_I[n] - \sum_{i=1}^Lv_i^{\mbox{\scp ext}}n_i
- \mu \sum_{i=1}^L n_i\right]
=0,
\eeq
where $\mu$ is a Lagrange multiplier and plays the role of a chemical
potential. We have then reduced the variational problem \rf{2.4} to \rf{2.6}
and \rf{2.7}. Unless we have a good approximation for $F_I[n]$, this
procedure is, of course, purely formal.

In order to proceed, we now suppose that we can define a simple Hamiltonian
$H_0^s$ acting on the same Hilbert space as \rf{2.2} and having the following
three properties. (i) It has the same global symmetries as $H_0$ in \rf{2.2}.
(ii) We are able to calculate its ground state even in the presence of external
inhomogeneous on-site potentials of the type given in \rf{2.3}, where the
Hamiltonian is given by
\beq \label{2.8}
H_s = H_0^s + \sum_{i=1}^L v_i^s \hat{n}_i,
\eeq
with $v_i^s$ ($i=1,\ldots,L$) arbitrary.
(iii) There exists a special choice of $v_i^s$ ($i=1,\ldots,L$) such that the 
ground state of \rf{2.8} has the same density $\{n_i\}$ as the ground state 
of the nontrivial interacting Hamiltonian \rf{2.2}. 
The auxiliary Hamiltonian $H_s$ is the Kohn-Sham Hamiltonian of DFT. It is 
not intended to approximate the complex problem posed by the Hamiltonian 
$H$, but to obtain the ground state densities and energies of $H$.\cite{kohnrmp,dftbook}

Property (i) is not rigorously necessary, but it simplifies our analysis
without too much restricting its generality. It allows us to consider $H_s$
on the sectors labelled by the eigenvalues $N=0,1,2,\ldots $ of the commuting
operator $\hat{N}$. Property (ii) is needed to make the formulation practical,
as it ensures that we can calculate the ground state energy and density
of the simpler Hamiltonian \rf{2.8}. Property (iii) is equivalent to a
v-representability assumption,\cite{dftbook} and is essential for the procedure to work.

Given these three properties, we can use the variational procedure
\rf{2.4}-\rf{2.7} to obtain the ground state energy from the minimization
\beq \label{2.9}
\delta \left[F_s[n] - \sum_{i=1}^Lv_i^s n_i -
\mu^s \sum_{i=1}^L n_i\right] =0,
\eeq
where, as in \rf{2.6},
\beq \label{2.10}
F_s[n] =
\mbox{min}_{|\Psi>_{n_1,\ldots,n_L}} <\Psi| H_0^s|\Psi>,
\eeq
and $\mu^s$ is a Lagrange multiplier. Comparison of \rf{2.9} and \rf{2.7}
implies that the density profile ($n_1,\ldots,n_L$) corresponding to the
ground state of the complicated Hamiltonian \rf{2.2} is the same as that
of the ground state of the simple Hamiltonian $H_s$, provide we choose
\beq \label{2.11}
\sum_{i=1}^L v_i^sn_i = \sum_{i=1}^Lv_i^{\mbox{\scp ext}}n_i + W[n] + (\mu
-\mu^s)N,
\eeq
where
\beq \label{2.12}
W[n] = F_I[n] - F_s[n].
\eeq
This implies
\beq \label{2.13}
v_i^s = v_i^{\mbox{\scp ext}} + \frac{\partial W[n]}{\partial n_i} + \mu -
\mu^s.
\eeq
For a given site $i$, the potential $v_i^s$ depends on the entire distribution
of local
densities $n_1,\ldots,n_L$. We have then to solve \rf{2.8} selfconsistently
for the ground state energy $E_s$ and the corresponding densities $\{n_1,
\ldots,n_L\}$. The difference $\mu - \mu^s$ in \rf{2.11} and \rf{2.13} is
just a harmless constant in the above selfconsistent procedure, since we are
always in a sector with a fixed value of $N$, and hence we may choose
$\mu = \mu^s$. As a consequence of \rf{2.11} and \rf{2.4}-\rf{2.6},  
the ground state energy of the complicated Hamiltonian \rf{2.2} is
\beq \label{2.15}
E[n] = E_s[n] - \sum_{i=1}^L \frac{\partial W[n]}{\partial  n_i} n_i +
W[n],
\eeq
which can be calculated once the density is known.

The success of this DFT formulation will depend on our ability to produce good
approximations for $W[n]$. In {\em ab initio} DFT, the functional $F_s$ is
just the kinetic energy $T_s$ of the noninteracting Hamiltonian $H_s$, and the 
functional $F_I[n]$ is decomposed as\cite{dftbook}
\beq
F_I[n]=T_s[n] + E_H[n] + E_{xc}[n],
\eeq
where $E_H$ is the Hartree energy (the mean-field approximation to the 
Coulomb interaction energy) and $E_{xc}$ the exchange-correlation
energy. $E_{xc}$ is unknown, and must be approximated. $E_H$ is known
explicitly, and $T_s$, although not known explicitly as a density
functional, is easily obtained from solving the Hamiltonian $H_s$. For general 
model Hamiltonians, the various terms in $F_I$ need not have the same
physical interpretation. We therefore simply write $F_I[n] = F_s[n]+W[n]$,
as above, and develop approximations for $W[n]$. 

If $F_I$ contains pieces that are known exactly, it is advantageous to 
treat these separately, and approximate only the remainder (e.g., $E_{xc}$ 
in the {\em ab initio} case). A typical case is the mean-field approximation 
to the interaction energy, leading to a Hartree-like contribution to $W$. 
Below we develop local approximations to $W[n]$, but in the models we are 
concerned with the Hartree term itself is local, so that nothing is gained
by giving it special treatment at this stage.

In order to turn the formulation into a practical tool, we require approximate 
expressions for $W[n]$ that keep the essential ingredients of the exact 
unknown functional. A general prescription for systematically producing such
approximations is to split the functional $W[n]$ into $L$ identical parts, 
according to
\beq \label{3.16}
W(n_1,\ldots,n_L) = \sum_{i=1}^L w_i
\eeq
where
\beq \label{3.16b}
w_i \equiv \frac{W(n_1,\ldots,n_L)}{L}, \; \; i=1,\ldots,L. 
\eeq
We now expand, for a given inhomogeneous distribution $n_1,\ldots,n_L$, 
each of the $L$ terms $w_i$ around a distinct homogeneous distribution,
\bea \label{3.17}
w_i& =& \frac{W(n_i,\ldots,n_i)}{L} +
\sum_{j=1}^L \alpha_j[n_i] (n_j-n_i) \nonumber \\
& +&
\sum_{j=1}^L\sum_{l=1}^L \beta_{j,l}[n_i] (n_j-n_i)(n_l-n_i) + \cdots.
\eea
The i'th term is thus expanded around a homogeneous distribution in which 
all sites have the density $n_i$ of the i'th site in the inhomogeneous 
distribution. Here $\alpha_j[n_i]$ and $\beta_{j,l}[n]$ depend on the local 
densities,
\bea \label{3.18}
\alpha_j[n_i] &=& \frac {1}{L} \frac{\partial W(n_1,\ldots,n_L)}{\partial
n_j}{\Bigg |}_{n_1=\cdots =n_L=n_i},  \nonumber \\
\beta_{j,l}[n_i] &=& \frac{1}{2L} \frac{\partial^2 W}{\partial n_i \partial
n_l}
{\Bigg |}_{n_1=\cdots=n_L=n_i}.
\eea
The first term in this expansion provides a local density approximation
(LDA) to the functionals. The second and higher-order terms yield non-local
contributions similar to those obtained in gradient expansions of the density
functional for Coulomb-interacting electronic systems.

Most of our analysis in the next sections is based on the LDA, i.e., the 
first term in \rf{3.17}
\beq \label{3.20}
W^{\mbox{\scp LDA}} (n_1,\ldots,n_L) = \sum_{i=1}^L \frac{W(n_i,n_i,\ldots,n_i)}{L}.
\eeq
The exact functional $W$ evaluated at an infinite and homogeneous density 
distribution $n_1=\cdots=n_L=n_i$, reduces to the functional of the infinite 
homogeneous system, depending only on the variable $n_i$, whose value per 
site we denote as $w^{\infty}(n)$. The LDA then assumes the more familiar form
\beq \label{3.21}
W^{\mbox{\scp LDA},\infty} (n_1,\ldots,n_L) = \sum_{i=1}^L w^{\infty}(n)|
_{n=n_i}.
\eeq
To compare this with the LDA in use in electronic-structure theory, recall
that there the Hartree term is nonlocal, and the local approximation is
only applied to the difference $E_{xc}=F_I-F_s-E_H$, whereas here we apply
it directly to $W=F_I-F_s$.

In this paper we explore the possibility to obtain the critical exponents
and conformal anomaly of critical Hamiltonians, by means of simple functionals
constructed following the above prescription. As a consequence of conformal
invariance,\cite{cardy} these quantities are obtained from the amplitudes
of the $O(1/L^2)$ finite-size corrections of the energies per site of the
quantum chains of size $L$. A finite-size quantum chain can still be
homogeneous, if all sites have the same spin, and periodic boundary
conditions are applied. Although the LDA becomes exact in the infinite
homogeneous system, in a finite homogeneous system it will have an additional 
error. If such error is of $O(1/L^2)$, we are going to obtain wrong critical 
exponents.

An improvement is obtained by using the  ground state energy per site
of a translationally invariant Hamiltonian with $L$ sites, instead of that
of the infinite system. In this case the functional depends on the size
$L$ of the system, and can be written as
\beq \label{3.22}
W^{\mbox{\scp LDA,L}} (n_1,\ldots,n_L) = \sum_{i=1}^Lw^{(L)}(n)|_{n=n_i},
\eeq
where $e^{(L)}(n_i)$ is the  ground state energy per site of the Hamiltonian
$H_0$. 

Note that this LDA-like expression is not, strictly speaking, a local 
functional, as it depends on the size of the system (i.e., the number of
elements in the set $\{n_1\ldots n_L\}$), which is a highly nonlocal
function of $n_i$. The resulting effective magnetic field $h_i^s$ at site $i$ 
does not depend only on the density at that site (as it does in local 
approximations), but on the number of sites. In applications 
reported below, this nonlocality endows the functional with superior
properties, compared to the usual LDA. In spite of this nonlocality, we 
call this latter functional the {\em finite-size LDA}, because of its
conceptual similarity with the usual (infinite-size) LDA, deriving from
the thermodynamic limit. A different type of nonlocality is explored in
Sec.~\ref{secV}.

Both local-density approximations, (\ref{3.21}) and (\ref{3.22}), as well as 
the gradient expansion (\ref{3.16})-(\ref{3.18}), are applicable to very
general classes of model Hamiltonians. In the remainder of this paper, we
employ concepts arising in the context of the renormalization group, 
in order to exemplify and test these approximations. The concept of
{\em universality classes} of models near {\em criticality} allows us to
consider one representative of critical quantum spin chains, and deduce 
from it results expected to be valid for the entire class. Specifically,
we use the LDAs \rf{3.21} and \rf{3.22} to obtain the energies of the 
XXZ chain and test the possibility of obtaining the conformal anomaly and 
critical exponents from LDA. The {\em conformal invariance} of the model 
allows us to obtain predictions for these quantities from energies
of finite chains. (Later on, in Sec.~\ref{secV}, conformal invariance
is used not only to extract information from functionals, but also to construct
nonlocal approximations.) For Kohn-Sham Hamiltonians that belong
to the same universality class as the interacting Hamiltonians, universality
guarantees that LDA critical exponents are given correctly. Nonuniversal
quantities, such as the sound velocity, are predicted wrongly by some forms 
of LDA, but a suitable rescaling of the Hamiltonian can be used to cure this 
defect.

\section{The XXZ quantum chain}\label{secIII}

Critical quantum chains belonging to the same universality class have the
same critical exponents and, in the case of conformally invariant systems,
the same conformal anomaly $c$. We expect that density functionals for
such systems will share the same general features.
In this section we consider the universality class comprising the $c=1$
conformally invariant quantum chains. These quantum chains exhibit a critical
line with continuously varying critical exponents. The underlying field theory
describing the long-distance physics of such models
is the Gaussian model (or Coulomb gas).\cite{gaussian} This field theory is
described by operators $\Phi_{n,m}$ composed of a spin-wave excitation index
$n$ and a "vortex" excitation of vorticity $m$. The anomalous dimensions of
these operators, related to the critical exponents of the quantum chain, are
given by
\beq \label{4.1}
x_{n,m} = n^2 x + \frac{m^2}{4x}, \;\;\; n,m=0,\pm1,\pm2,\ldots .
\eeq
The parameter $x$ varies continuously along the critical line and its value
depends on the interactions entering the particular model.\cite{gaussian-note}
Examples of models exhibiting such critical behavior are the anisotropic
spin-$S$ Heisenberg model,\cite{alc-moreo} the Ashkin-Teller model,\cite{ab2} 
the anisotropic triplet spin model introduced in Ref.~\onlinecite{alc-barber}, 
the ferromagnetic spin-$S$ Babujian-Taktajan models,\cite{alc-martins} the 
anisotropic biquadratic spin $1$ chain,\cite{biqua} etc. 

The most studied model in this family of
$c=1$ models is the anisotropic spin-$\frac{1}{2}$ Heisenberg chain, also
known as the XXZ chain, whose Hamiltonian is given by
\beq \label{4.2}
H^{\mbox{\scp XXZ}}(\Delta) = -\frac{1}{2} \sum_{j=1}^L
(\sigma_j^x\sigma_{j+1}^x + \sigma_j^y\sigma_{j+1}^y + \Delta
\sigma_j^z\sigma_{j+1}^z).
\eeq
Here $\vec{\sigma} \equiv (\sigma^x,\sigma^y,\sigma^z)$ are Pauli matrices,
$L$ is the lattice size and $\Delta$ is the anisotropy, which in the DFT
framework plays the role of a spin-spin interaction. This Hamiltonian is
exactly integrable through the Bethe ansatz.\cite{yang-yang} Its exact solution
shows that, in the bulk limit $L \rightarrow \infty$, the model is critical
(gapless) for anisotropies $-1\leq \Delta = -\cos \gamma \leq 1$. As a
consequence of the conformal invariance of the model, the critical exponents
are exactly given by Eq~\rf{4.1},\cite{ab2,hamer} with the model dependent
parameter $x = x_{\Delta}$ given by
\beq \label{4.3}
x_{\Delta} = \frac{\pi-\gamma}{2\pi}, \; \; \; \Delta = - \cos \gamma,
\; \; 0 \leq \gamma \leq \pi.
\eeq
The XXZ quantum chain is then a natural candidate to develop approximate
functionals for DFT applications to critical models in the $c=1$
universality class. The exactly known finite-size corrections of the 
model\cite{ab2} will allow us to test the general ideas presented in Sec. 
\ref{secII}.

As in Sec. \ref{secIIB}, we consider a non integrable generalization of 
\rf{4.2}
where besides the intersite spin-spin interactions we also include the
on-site interactions of the spins with an inhomogeneous site-dependent
external magnetic field $h_i^{\mbox{\scp ext}}$
($i=1,\ldots,L$), namely,
\bea \label{4.4}
H^{\mbox{\scp XXZ}}(\Delta, \{h_i\}) = H_0^{\mbox{\scp XXZ}}(\Delta) +
V^{\mbox{\scp ext}} 
\nonumber \\
=\frac{1}{2} \sum_{j=1}^L
(\sigma_j^x\sigma_{j+1}^x + \sigma_j^y\sigma_{j+1}^y + \Delta
\sigma_j^z\sigma_{j+1}^z)
\nonumber \\
- \frac{1}{2}\sum_{j=1}^Lh_j^{\mbox{\scp ext}}(\sigma_j^z +1),
\eea
where we added a convenient constant. In this paper we study the
Hamiltonian \rf{4.4} with several types of boundary conditions. For
periodic ($p=1$) and open boundary conditions ($p=0$) we have
\beq \label{4.5}
\vec{\sigma}_{L+1}= p \vec{\sigma}_1, \;\;\; p =0,1,
\eeq
whereas for twisted boundary conditions
\beq \label{4.6}
{\sigma^{\pm}}_{L+1}= e^{\pm i \phi} {\sigma}^{\pm}_1, \;\;\;
\sigma_{L+1}^z = \sigma_1^z,
\eeq
where 
$0\leq \phi \leq 2\pi$ and $\sigma^{\pm} = \frac{1}{2}(\sigma^x \pm i\sigma^y)$
are the usual raising and lowering $SU(2)$ spin operators.

The inhomogeneous Hamiltonian \rf{4.4} is suitable for the DFT approach
presented in Sec. \ref{secII}. It has a $U(1)$ symmetry, and it is easy to 
obtain a simple Hamiltonian $H^s$, as in \rf{2.8}, with the properties 
(i)-(iii) discussed in Sec. \ref{secII}. The Hamiltonian
\rf{4.4} commutes with the global charge
\beq \label{4.7}
\hat{N} = \sum_{i=1}^L \hat{n}_i, \; \; \; \hat{n}_i = \frac{\sigma_i^z+1}{2},
\; \;
i =1,\ldots, L,
\eeq
which gives the total number of up spins in the $\sigma^z$-basis. By comparing
\rf{2.2}, \rf{2.3} and \rf{4.4} we identify
\beq \label{4.8}
V^{\mbox{\scp ext}} = \sum_{i=1}^L v_i^{\mbox{\scp ext}} \hat{n}_i, \;\;\;
v_i^{\mbox{\scp ext}} = -h_i, \;\;\;\
i=1,\ldots,L.
\eeq
The simple Hamiltonian that plays the role of the Kohn-Sham Hamiltonian is
obtained by setting $\Delta =0$ in \rf{4.4}, i.e.,
\beq \label{4.9}
H_s(\{h_i^s\}) = H^{\mbox{\scp XXZ}}(\Delta=0,\{h_i^s\}) =
H^{\mbox{\scp XXZ}}(0) - \sum_{i=1}^Lh_i^s\hat{n}_i.
\eeq
This simple Hamiltonian is the well known XY model in the presence of
site-dependent magnetic fields $h_i^s$ ($i=1,\ldots,L$). Through a
Jordan-Wigner transformation\cite{jordan-wigner} 
$H_s$ is tranformed into a Hamiltonian describing 
$L$ non-interacting spinless fermions. The $2^L$ eigenenergies of
$H_s$ are given by arbitrary combinations of free fermion energies,
given in terms of the eigenvalues of a $L\times L$ matrix, whose elements
depend on the boundary condition in \rf{4.9} and the values of the magnetic
fields $h_i^s$ ($i=1,\ldots,L$). This implies that with reasonable computing
efforts we can diagonalize $H_s$ for lattice sizes up to $L=L_{\mbox{\scp max}}
\approx 5000$. This is certainly much more we could reach for the
$2^L \times 2^L$ interacting Hamiltonian \rf{4.4}.

The inhomogeneous magnetic field $h_i^s$ ($i=1,\ldots,L$) are fixed, as in 
\rf{2.13}, by imposing that the ground state eigenfunction of \rf{4.9} and
\rf{4.4}, on a given sector with fixed number $N$ of up spins, share the same
density distribution $n_i$ ($i=1,\ldots,L$) of up spins:
\beq \label{4.10}
h_i^s = h_i^{\mbox{\scp ext}} - \frac{\partial W[n]}{\partial n}, \;\;\;\ i=1,
\ldots,L.
\eeq
As in \rf{4.12}, $W[n]$ is given by the difference of the functionals
of the interacting \rf{2.6} and non-interacting \rf{2.10} models. Since in the
present case $H_s$ is obtained from $H$ by setting $\Delta =0$, the
corresponding functionals are related and we can write
\beq \label{4.11}
W[n] = F_{\Delta}[n] - F_{\Delta=0}[n],
\eeq
where $F_{\Delta}$ is the functional obtained by using in \rf{2.6} the
Hamiltonian
$H^{\mbox{\scp XXZ}}(\Delta)$ given in \rf{4.2}, i.e.,
\beq \label{4.12}
F_{\Delta}[n] =
\mbox{min}_{|\Psi>_{n_1,\ldots,n_L}} <\Psi| H^{\mbox{\scp XXZ}}(\Delta)|\Psi>.
\eeq
The density profile $n^0 = (n_1^0,\ldots,n_L^0)$ of the ground state of
$H^{\mbox{\scp XXZ}} (\Delta,\{h_i\})$ restricted to the sector with $N$ up
spins,
is obtained by solving self-consistently \rf{4.9} with \rf{4.10}. From
\rf{2.15} the ground state energy of the inhomogeneous interacting
model \rf{4.4} is given by
\beq \label{4.13}
E_N = E[n^0] = E_s(\{h_i^s\}) -
\sum_{i=1}^L \frac{\partial W[n^0]}{\partial n_i^0} n_i^0 + W[n^0].
\eeq
Here $E_s(\{h_i^s\})$ is the ground state energy of the free fermion XY
Hamiltonian $H^{\mbox{\scp XXZ}}(\Delta=0,\{h_i^s\})$, with
$N=\sum_{i=1}^L n_i^0$ up spins.

All difficulties in obtaining the energies $E_N$ in \rf{4.13} now reduce to
the derivation of the functional $W_{\Delta}[n]$ in \rf{4.11}. Following the
approximations considered in Sec. \ref{secII}, we next consider the XXZ chain 
with distinct boundary conditions and inhomogeneities.

\section{Local approximations for the XXZ chain}\label{secIV}

In the absence of external fields $h_i$, the only source of inhomogeneity 
(i.e., breaking of translational invariance) are the boundaries. In this 
section we consider the Hamiltonian \rf{4.2} on finite chains, for
periodic, twisted and open boundary conditions.

\subsection{Periodic boundary conditions}\label{secIVA}

In this case, beyond the conservation of the total number of up spins, the 
Hamiltonian \rf{4.2} is also translationally invariant. As a consequence, 
we can separate the Hilbert space associated with the interacting model 
$H^{\mbox{\scp XXZ}}(\Delta)$, as well as that of the auxiliary Kohn-Sham 
Hamiltonian $H^{\mbox{\scp XXZ}} (\Delta=0)$, 
into block-disjoint sectors labelled by the density of up spins 
$n = \frac{N}{L}$, ($n=0,\frac{1}{L},\frac{2}{L},\ldots,1$) and 
momentum $p = \frac{2\pi}{L}j$ ($j=0,1,\ldots,L-1$). In each of these sectors 
we can apply the DFT-LDA procedure of section \ref{secII} to calculate the 
lowest energy $E_0(L,n,p)$.

Let $|\phi>$ be any vector on the arbitrary sector ($n,p$), i.e., 
$\sum_{i=1}^L {\hat n}_i/L |\phi> = n|\phi>$ and $\hat{T} |\phi> = e^{ip}|
\phi>$, where $\hat{T}$ is the translation operator. It is simple to show that 
\beq \label{5.1}
n_i = \frac{<\phi|\hat{n}_i|\phi>}{<\phi|\phi>} = 
\frac{\sum_{i=1}^L n_i}{L} = \frac{N}{L}=n, \;\;\; i=1,\ldots,L. 
\eeq
This implies that any eigenvector of $H^{\mbox{\scp XXZ}}(\Delta)$ has a 
uniform distribution of densities $n_i=n$ ($i=1,\ldots,L$). 

The functional $F_{\Delta}[n,p] = F_{\Delta}(n,p)$, for general values of 
$\Delta$, is in the present case a simple function of the two variables 
$n=N/L$ and $p$. 
Consequently the local magnetic fields $\{h_i^s\}$ in the auxiliary Hamiltonian 
\rf{4.9} are site independent:
\beq \label{5.2}
h_i^s = -\frac{d W_{\Delta}(n,p)}{d n} = 
-\frac{d } {d n} (F_{\Delta}(n,p) - F_{0}(n,p)) \equiv h^s.
\eeq
$H_s(\{h_i^s\})$, in this case, is just the XY model in the presence of a 
uniform magnetic field $h_s$. The lowest energy of this auxiliary Hamiltonian, 
in the sector $(n,p)$, is given by
\beq \label{5.3}
E_s(h^s) = L e^{L,\mbox{\scp per}}_{\Delta=0}(n,p) - L h^s n,
\eeq
where $e^{L,\mbox{\scp per}}_{\Delta=0}(n,p)=e^{L,\mbox{\scp per}}_0(n,p)$ 
is the lowest eigenenergy per site in the 
sector $(n,p)$ of the XY model in the absence of external magnetic fields. 

In order to obtain the lowest eigenenergies of the interacting model on the 
sector $(n,p)$, from \rf{4.13}, it is necessary to know the functional 
$W_{\Delta}(n,p)$. Following the discussions of Sec. \ref{secIII}, we use the 
local approximations \rf{3.21} and \rf{3.22} for this functional. In the 
present case they are given by 
\beq \label{5.4}
W_{\Delta}^{\mbox{\scp LDA},L}(n,p) = 
\sum_{i=1}^L w_{\Delta}^{L,\mbox{\scp per}}(n,p) = 
L w_{\Delta}^{L,\mbox{\scp per}}(n,p),
\eeq
\beq \label{5.5}
W_{\Delta}^{\mbox{\scp LDA},\infty}(n,p) 
= \sum_{i=1}^L w_{\Delta}^{\infty}(n,p) = 
L w_{\Delta}^{\infty}(n,p).
\eeq
Here $w_{\Delta}^{L,\mbox{\scp per}}(n,p)=e_{\Delta}^{L,\mbox{\scp per}}(n,p)-e_0^{L,\mbox{\scp per}}(n,p)$ and $w_{\Delta}^{\infty}(n,p) =
e_{\Delta}^{\infty}(n,p) - e_0^{\infty}(n,p)$ 
are the interaction energies per site, and 
$e_{\Delta}^{L,\mbox{\scp per}}(n,p)$ and 
$e_{\Delta}^{\infty}(n,p)$ represent the lowest total 
energy per site in the sector $(n,p)$ of the periodic Hamiltonian 
$H^{\mbox{\scp XXZ}}(\Delta,p)$ with $L$ sites and with 
an infinite number of sites, respectively. The Hamiltonian 
$H^{\mbox{\scp XXZ}}(\Delta,0)$, with periodic boundaries, is exactly 
integrable through the Bethe ansatz.\cite{yang-yang} The eigenenergies for 
lattice size $L$, are obtained by solving a set of $N=nL$ coupled non-linear 
equations. This can be done,\cite{ab2} at least for the lower eigenenergies 
of sectors $(n,p)$, either for lattice sizes up to $L \approx 1000$ or in 
the bulk limit ($L\rightarrow \infty$).

Suppose, for the moment, that we have evaluated exactly 
$e_{\Delta}^{L,\mbox{\scp 
per}}(n,p)$ and the LDA functional \rf{5.4} and \rf{5.5}. The lowest 
eigenenergy, in the sector $(n,p)$, is obtained by inserting \rf{5.3} 
in \rf{4.13} and using one of the approximations \rf{5.4} or \rf{5.5}. 
From the finite-size LDA functional \rf{5.4} we obtain
\bea \label{5.6}
E_{\Delta}^{\mbox{\scp LDA},L}(L,n,p) =
L e^{L,\mbox{\scp per}}_0(n,p) - L h^s n 
\nonumber \\
- \sum_{i=1}^Ln \frac{\partial W_{\Delta}(n,p)}{\partial n} + W_{\Delta}(n,p).
\eea
The relation \rf{5.2} gives, in this case, the exact result
\beq \label{5.7}
E_0^{\mbox{\scp LDA},L} (L,n,p) = L e_{\Delta}^{L,\mbox{\scp per}} (n,p).
\eeq
This result is expected, since the density distribution of up spins for any 
of the eigenlevels is spatially homogeneous, and we only need to keep the 
first term in the expansion \rf{3.16}-\rf{3.18}. The conformal anomaly and 
critical exponents that are calculated from the leading finite-size corrections 
of the eigenenergies, are then obtained exactly. 

This is not the case for the LDA functional \rf{5.5}, which derives from the 
infinite chain. If we use \rf{5.5} and \rf{5.3} in \rf{4.13}, we obtain
\bea \label{5.8} 
E_{\Delta}^{\mbox{\scp LDA},\infty}(L,n,p) =
L e_0^{L,\mbox{\scp per}}(n,p) - L h^s n 
\nonumber \\
-\sum_{i=1}^L[\frac{\partial}{\partial n}(e_{\Delta}^{\infty}
(n,p) - e_0^{\infty}(n,p))] n \nonumber \\
+ L(e_{\Delta}^{\infty}(n,p) - 
e_0^{\infty}(n,p)).
\eea
Equation \rf{5.2} gives
\bea \label{5.9}
E_{\Delta}^{\mbox{\scp LDA},\infty} (L,n,p) =
\nonumber \\
L[ e_{\Delta}^{\infty}(n,p)
+ e_0^{L,\mbox{\scp per}}(n,p)-e_{0}^{\infty}(n,p)].
\eea
This result is certainly not exact for all orders of $1/L$, unlike that 
obtained from the finite-size LDA functional \rf{5.4}.
However, in typical applications of LDA in {\em ab initio} electronic-structure
theory, as well as to the Hubbard and the Heisenberg model, one employs 
an LDA functional deriving from the corresponding infinite system, for 
any calculation on finite-size systems. In the present work, this approach
is represented by the infinite-size LDA functional \rf{5.5}. It is therefore 
interesting to evaluate the leading finite-size corrections of \rf{5.9},
obtained from \rf{5.5}, to see if we can, from this more standard LDA,
still obtain exact results for the conformal anomaly and critical exponents.

The ground state of the XXZ chain has zero momentum. It belongs to the 
half-filled sector with $n=\frac{N}{L}=\frac{1}{2}$, for even values of $L$, 
and to the sector $n= \frac{1}{2} \pm \frac{1}{L}$ in the case of odd values 
of $L$. Let us restrict ourselves to the case where $L$ is even. The asymptotic
behavior of the ground state energy is exactly known\cite{ab2,hamer}
\bea\label{5.10}
E_{\Delta}^{\mbox{\scp per}} (L; \frac{1}{2},0) = 
Le_{\Delta}^{L,\mbox{\scp per}}(\frac{1}{2},0) 
\nonumber \\
=Le_{\Delta}^{\infty}(\frac{1}{2},0) 
- \frac{\pi v_{\Delta}c_{\Delta}}{6L} 
+ o(\frac{1}{L}),
\eea
where $c_{\Delta} =1$ is the conformal anomaly and 
\beq \label{5.11}
v_{\Delta} = \frac{\pi\sin \gamma}{\gamma}, \; \;\; -1\leq \Delta =-\cos 
\gamma < 1, 
\eeq
is the $\Delta$-dependent sound velocity. Using \rf{5.10} with $\Delta=0$ in 
\rf{5.9} we obtain the LDA prediction for the asymptotic behavior of the 
ground state energy
\beq \label{5.12}
E_{\Delta}^{\mbox{\scp LDA},\infty}(L;\frac{1}{2},0) = 
Le_{\Delta}^{\infty}(\frac{1}{2},0) - \pi c_{\Delta}v_0/6L + 
o(\frac{1}{L}),  
\eeq
where $v_0 = v_{\Delta =0} = 2.$

Upon comparing \rf{5.12} with \rf{5.10}, we see that the LDA \rf{5.5} yields an 
incorrect amplitude for the $O(1/L)$ term. The sound velocity predicted is not 
given by \rf{5.11} but by its value at $\Delta =0$ ($\gamma =\frac{\pi}{2}$). 
This means that the LDA \rf{5.5} does not allow an exact prediction of the 
conformal anomaly. 

The critical exponents, as a consequence of the conformal invariance of the 
infinite system, are obtained from the finite-size corrections of the energy 
gaps of the quantum chain. The lowest eigenenergies on the sectors with 
$N=\frac{L}{2} + \nu$ ($\nu = 0, \pm 1, \pm 2, \ldots$) up spins (density 
$n =\frac{N}{L} = \frac{1} {2} + \frac{\nu}{L}$) are zero-momentum states. 
The asymptotic behavior of these energies is exactly known\cite{ab2,hamer} 
\bea \label{5.13}
E_{\Delta}(L;\frac{1}{2} +\frac{\nu}{L},0) = 
Le_{\Delta}^{L,\mbox{\scp per}}(\frac{1}{2} +\frac{\nu}{L},0) = 
\nonumber \\
Le_{\Delta}^{\infty}(\frac{1}{2} +\frac{\nu}{L},0) -\frac{\pi 
v_{\Delta} 
c_{\Delta}}{6L} +
\frac{2\pi}{L} v_{\Delta} x_{\Delta} \nu^2 + o(\frac{1}{L}),
\eea
where $x_{\Delta}$ is given by \rf{4.3}. The energy gaps related to the 
eigenenergies \rf{5.13} then give us the parameter $x_{\Delta}$ that 
characterizes the particular $c=1$ conformal theory:
\beq \label{5.14}
G_L^{\nu} = E_{\Delta}(L;\frac{1}{2} +\frac{\nu}{L},0) 
- E_{\Delta}(L;\frac{1}{2} ,0) = \frac{2 \pi}{L} v_{\Delta}x_{\Delta} \nu^2 
+ o(\frac{1}{L}), 
\eeq
where $\nu = \pm 1 \pm 2, \ldots$.

The gaps \rf{5.14} predicted by the LDA functional \rf{5.5} are obtained 
from  \rf{5.9} 
\bea \label{5.15}
G_L^{{\mbox \scp LDA},\nu} =
\nonumber \\ 
E_{\Delta}^{\mbox{\scp LDA},\infty} (L; \frac{1}{2} + \frac{\nu}{L}, 0)  
-E_{\Delta}^{\mbox{\scp LDA},\infty} (L; \frac{1}{2} , 0)  
\nonumber \\
=L(e_{0}^{L,\mbox{\scp per}} (\frac{1}{2} + \frac{\nu}{L}, 0)  
-e_{0}^{L,\mbox{\scp per}} ( \frac{1}{2} , 0) ) \nonumber \\ 
+ L(e_{\Delta}^{\infty} ( \frac{1}{2} + \frac{\nu}{L}, 0)  
-e_{\Delta}^{\infty} ( \frac{1}{2} , 0) ) 
\nonumber \\
-L(e_{0}^{\infty} ( \frac{1}{2} + \frac{\nu}{L}, 0)  
-e_{0}^{\infty} ( \frac{1}{2} , 0) ).
\eea
The leading behavior, as $L\rightarrow \infty$, of the energies in the first 
parenthesis of the last expression is given by \rf{5.13} with $\Delta =0$. 
In order to find the leading finite-size corrections of the gaps \rf{5.15}, we 
need to calculate those corrections for 
$e_{\Delta}^{\infty}(\frac{1}{2}+\frac{\nu}{L}, 0)$.  This is the lowest 
eigenenergy of the infinite system with finite density $ n = \frac{1}{2} + 
\frac{\nu}{L} $ of up spins and zero momentum. It is important to stress that 
$\nu/L$ is small but finite. From \rf{5.13} we have the exact behavior 
\bea \label{5.16}
e_{\Delta}^{\infty} (\frac{1}{2} +\frac{\nu}{L},0) =  
\lim _{L'\to \infty} E_{\Delta}(L',\frac{1}{2}+\frac{\nu}{L},0)/L'  \nonumber \\
= e_{\Delta}^{\infty}(\frac{1}{2},0) +\frac{2\pi}{L'^2} 
(\frac{L'\nu}{L})^2 x_{\Delta}v_{\Delta} + o(\frac{1}{L^2}) \nonumber \\
= e_{\Delta}^{\infty}(\frac{1}{2},0) +\frac{2\pi 
v_{\Delta}x_{\Delta}}{L^2} \nu^2 + o(\frac{1}{L^2}). 
\eea
By substituting \rf{5.16} and \rf{5.13} in \rf{5.15} we find
\beq \label{5.17}
G_L^{\mbox{\scp LDA},\nu} = \frac{2 \pi}{L} v_{\Delta}x_{\Delta}\nu^2 + 
o(\frac{1}{L}), \;\; 
\nu =\pm1,\pm2,\ldots,
\eeq
which reproduces the exact result \rf{5.14}. This means that although the 
infinite-size LDA \rf{5.5} does not give an exact value for the conformal 
anomaly $c$, it produces exact results for the critical exponents 
$x_{\Delta}\nu^2$. In order to obtain the exact results  \rf{5.17}, it is 
crucial to use the exact result for the energy per site of the homogeneous
model, $e_{\Delta}^{\infty} (n,p)$, which can be obtained 
from the Bethe ansatz solution of the model.\cite{yang-yang} 
For completeness we give, in the appendix, the 
relevant integral equations that produce exact values for 
$e_{\Delta}^{\infty} (n,p)$. In Sec. \ref{secV} we introduce 
an analytical approximation for the functional \rf{5.5} that contains all 
relevant ingredients to reproduce the exact results of the critical exponents 
given in \rf{5.17}.

Let us now consider the LDA predictions for the mass gap amplitudes related to 
the sectors with momentum $p$. The energy-momentum dispersion relation gives 
\beq \label{5.18}
G_L^p=
L(e_{\Delta}^{L,\mbox{\scp per}}(L;\frac{1}{2},p) - e_{\Delta}^{L,\mbox{ \scp 
per}}(L;\frac{1}{2},0)) 
= v_{\Delta} p, 
\eeq
where $v_{\Delta}$ is the sound velocity \rf{5.11} and $p = \frac{2\pi}{L}j$ 
($j=0,1,\ldots$). From this last expression we also obtain 
\bea \label{5.19}  
e_{\Delta}^{\infty}(\frac{1}{2},p) -
e_{\Delta}^{\infty}(\frac{1}{2},0) =
\nonumber \\
= \lim_{L' \to \infty} \frac{1}{L'} [ L'(
e_{\Delta}^{L',\mbox{\scp per}}(L';\frac{1}{2},p) - 
e_{\Delta}^{L',\mbox{\scp per}}(L';\frac{1}{2},0))]  \nonumber \\
= 
\lim_{L' \to \infty} v_{\Delta}p/L' = 0.
\eea
From \rf{5.9}, \rf{5.18} and \rf{5.19} we obtain the LDA prediction for 
the momentum mass gap 
\beq \label{5.20}
G_L^{\mbox{\scp LDA},p} = v_0p, \; \; \; v_0=2. 
\eeq

Again, as in \rf{5.12}, we obtain the same incorrect prediction for the 
sound velocity of the model, i.e., the value of the non interacting XY model. 
However, the sound velocity is a model-dependent quantity. Its value changes 
if the Hamiltonian is multiplied by an arbitrary positive constant, without 
changing the ratios of mass gaps that are related to the critical exponents. 
On the other hand, the underlying conformal field theory governing the 
critical line of the model should have a fixed sound velocity. This 
expectation is bourn out by the XXZ quantum chain if we consider in \rf{4.2} 
the rescaled Hamiltonian $H^{\mbox{XXZ}}(\Delta)/v_{\Delta}$, where 
$v_{\Delta}$ is given by \rf{5.11}. In this case the 
sound velocity becomes unity for any value of $\Delta$ and we obtain, 
even from the infinite-size LDA \rf{5.5}, exact results for the sound velocity 
($v_{\Delta} =1$), conformal anomaly ($c=1$) and critical exponents 
($x_{\Delta}\nu^2$).

Before finishing our discussion of the periodic case, it is also 
interesting to consider the mass-gap amplitudes related to the special sector 
with momentum $p = \pi$. The numerical solution of the Bethe ansatz equations 
for finite chains,\cite{ab2} and the conformal invariance of the infinite 
system predict, from the finite-size corrections related to this gap, the 
critical exponent $x_{0,1} = 1/4x_{\Delta}$, where $x_{\Delta}$ is given by 
\rf{4.3}. As before, the infinite-size LDA \rf{5.5} reproduces this exact 
result. 

\subsection{Twisted boundary conditions}\label{secIVB}

In this case, the boundary condition \rf{4.6} is specified by the angle 
$0\leq \phi <2\pi$, the periodic case corresponding to $\phi=0$. The total 
density $n$ of up spins remains a good quantum number. Moreover, the 
quantum chain also displays a generalized translation invariance for arbitrary 
values of $\phi$ (see Ref.~\onlinecite{mome} for the proper definition of 
translations on the lattice). As a consequence of this invariance, as in
Eq.~\rf{5.11}, the density obtained from any given eigenstate of the 
Hamiltonian is homogeneous, i.e., 
$n_i = N/L = n$ ($i=1,\ldots,L$). For the sake of simplicity, in the 
following we restrict ourselves to states with zero generalized momentum. 
The LDA functionals, obtained from \rf{3.20} and \rf{3.21}, are generalizations
of \rf{5.4} and \rf{5.5}, and now take the form
\beq \label{5.21} 
W_{\Delta}^{\mbox{\scp LDA},\phi}(n) = L w_{\Delta}^{L,\phi}(n),
\eeq
and
\beq \label{5.22}
W_{\Delta}^{\mbox{\scp LDA},\infty}(n) = L w_{\Delta}^{\infty}(n).
\eeq
Here $w_{\Delta}^{L,\phi}(n) = e_{\Delta}^{L,\phi}(n) - e_0^{L,\phi}(n)$ and 
$w_{\Delta}^{\infty}(n)=e_{\Delta}^{\infty}(n)-e_0^{\infty}(n)$ 
are the interaction energies, per site, in the eigensector with density $n$ of 
up spins and lowest energy $ e_{\Delta}^{L,\phi}(n)$ or 
$e_{\Delta}^{\infty}(n)$ of the XXZ chain with twisted boundary 
conditions specified by the angle $\phi$.

Let us restrict ourselves to the case where $L$ is even. The ground state 
energy belongs to the eigensector labelled by the density $n=1/2$ of up 
spins. The Bethe ansatz gives the asymptotic behavior of the ground state 
energy for $L \rightarrow \infty$,
\beq \label{5.23}
E_{\Delta}^{\phi}(L,\frac{1}{2}) = Le_{\Delta}^{\infty} (\frac{1}{2}) 
-\frac{\pi v_{\Delta}}{6L} \tilde{c}_{\Delta} + o(\frac{1}{L}).
\eeq
Here $e_{\Delta}^{\infty}(\frac{1}{2})$ is the ground state energy, per 
site, in the bulk limit, and 
\beq \label{5.24} 
\tilde{c}_{\Delta} = 1- 12x_{0,\frac{\phi}{\pi}} = 
1 - 12 (\frac{\phi}{\pi})^2 /4x_{\Delta},
\eeq
is the conformal anomaly of an effective model described by the XXZ quantum 
chain with twisted boundary condition specified by $\phi$.\cite{ab2}
The parameters $x_{n,m}$, $v_{\Delta}$ and $x_{\Delta}$ in Eqs.~\rf{5.23}
-\rf{5.24} are given by Eqs.~\rf{4.1},\rf{5.11} and \rf{4.3}, respectively. 
The exact results\cite{ab2} for the leading behavior of the mass gaps of the 
eigensectors with density 
$n=\frac{1}{2} +\frac{\nu}{L}$ ($\nu = \pm 1, \pm 2, \ldots$) of up spins yield
\beq \label{5.25}
G_L^{\nu,\phi} = 
E_{\Delta}^{\phi} (L,\frac{1}{2} +\frac{\nu}{L}) -
E_{\Delta}^{\phi} (L,\frac{1}{2} ) =
\frac{2 \pi}{L} v_{\Delta}x_{\Delta}\nu^2 + o(\frac{1}{L}), 
\eeq
independently of the boundary angle $\phi$.

As in the periodic case, we now test the possibility to recover the 
asymptotic behavior specified by Eqs. \rf{5.23} and \rf{5.25} from the 
LDA functionals \rf{5.21} and \rf{5.22}. Similarly to the case $\phi=0$ 
(periodic boundary conditions), if we use the finite-size LDA \rf{5.21} 
we obtain the exact result 
\beq \label{5.26} 
E_0^{\mbox{\scp LDA},L,\infty}(L,n) = L e_{\Delta}^{L,\phi}(n),
\eeq
and thus the correct value for the effective conformal anomaly 
$\tilde{c}_{\Delta}$ and the critical exponent $x_{\Delta}\nu^2$. This 
happens because in this case \rf{5.21} is exact. For any eigenstate of the 
Hamiltonian the density is homogeneous ($n_1=\cdots =n_L$) and the only
nonvanishing term in the expansion \rf{3.17} is the first one.

Let us now consider the infinite-size LDA functional \rf{5.22}. The 
expressions for the LDA eigenenergies, which for the periodic case ($\phi=0$) 
were given by \rf{5.9}, are now replaced by 
\beq \label{5.27} 
E_{\Delta}^{\mbox{\scp LDA},\infty} (n) = 
L[e_{\Delta}^{\infty}(n) 
+ e_{0}^{L,\phi}(n) 
- e_{0}^{\infty}(n)] .
\eeq
By evaluating Eq.~\rf{5.23} for $\Delta=0$ we obtain from Eq.~(\ref{5.27})
the asymptotic behavior of the ground state energy, as $L\rightarrow \infty$,
\beq \label{5.28}
E_{\Delta}^{\mbox{\scp LDA},\infty}(\frac{1}{2}) = L e_{\Delta}^{\infty,
\phi} (\frac{1}{2},0) - 
\pi \tilde{c}_{\Delta}v_0/6L + o(1/L).
\eeq
 Comparing \rf{5.28} with \rf{5.23} we obtain, as in the $\phi=0$ case, 
the sound velocity $v_0 =2$ of the XY model instead of the correct 
$v_{\Delta}$, given in \rf{5.11}.

As in the periodic case, in order to evaluate the critical exponents, we need 
the asymptotic large-$L$ behavior of $e_{\Delta}^{\infty}(\frac{1}
{2}+\frac{\nu}{L})$, with $\frac{\nu}{L}$ small but finite 
($\nu = \pm 1, \pm 2 , \ldots $). This is obtained 
from \rf{5.25} in a similar way as we got \rf{5.16} from \rf{5.13}:
\beq \label{5.29}
e_{\Delta}^{\infty}(\frac{1}{2}+\frac{\nu}{L}) =
e_{\Delta}^{\infty}(\frac{1}{2}) + 2\pi v_{\Delta}x_{\Delta} \nu^2/L^2 
+ o(1/L^2).
\eeq
By inserting the LDA energies \rf{5.27} in \rf{5.25}, and using \rf{5.29}, we 
obtain 
\beq \label{5.30}
G_L^{\mbox{\scp LDA},\nu,\infty} = \frac{2\pi}{L} v_{\Delta}x_{\Delta} \nu^2 + 
o(1/L).
\eeq
Thus, as in the periodic case, the infinite-size LDA \rf{5.22} gives exact 
results for the critical exponents, although it is not the exact functional
for the finite-size chain.

The wrong estimate obtained for the sound velocity is a 
consequence of the distinct sound velocities $v_{\Delta}$ and $v_0=2$ of the 
interacting and non-interacting Hamiltonians. If we consider the rescaled 
Hamiltonian $H^{\mbox{\scp XXZ}}(\Delta)/v_{\Delta}$, similar to what we did 
in the periodic case, the sound velocity is now unity for both Hamiltonians, 
and the prediction from the LDA \rf{5.22} is now exact both for the conformal 
anomaly and critical exponents.

\subsection{Open boundary conditions}\label{secIVC}

For open boundary conditions, the quantum chain \rf{4.2} still conserves 
the total number $N$ of up spins, but now is not translationally invariant. 
In general, the eigenfunctions belonging to the eigensectors with total 
density $n=N/L$ of up spins produce inhomogeneous distributions 
of local densities $n_1,\ldots,n_L$. Let us restrict ourselves, in the 
following, to cases where the lattice size $L$ is an even number.

The Hamiltonian \rf{4.2} with open boundaries commutes with the 
spin reversal and total $z$-magnetization operators,
\beq \label{5.31}
\hat{R} = \prod_{i=1}^L \sigma_i^x, \;\;\;\;\; S^z = \sum_{i=1}^L 
\sigma_i^z.
\eeq
When restricted to the sector where $S^z =0$, these operators also commute 
with each other, $[\hat{R},\hat{S}]=0$. This implies that in the eigensector 
with density $n=N/L=1/2$ of up spins, the parity under spin reversal 
is also a good quantum number. It is then simple to show that an arbitrary 
eigenfunction $|\Psi>$, on this sector, produces a homogeneous local 
density, namely,
\beq \label{5.31-b}
n_i = <\Psi|(\sigma_i^z +1)/2|\Psi>/<\Psi|\Psi> = \frac{1}{2}, \;\;\;\ 
i =1,\ldots,L.
\eeq
Here we will only consider this case, whose analysis is similar to that
presented for periodic and twisted boundary conditions. The general case,
in which the density is spatially inhomogeneous, is treated by means of
an improved functional in Sec.~\ref{secVB}.

The ground state of \rf{4.2} belongs to the special sector where 
$n=N/L=1/2$. The XXZ quantum chain with open boundaries is also exactly 
integrable through the Bethe ansatz.\cite{ab3q}
The exact asymptotic behavior of the ground state energy is known\cite{ab3q} 
to be
\bea \label{5.32}
E_{\Delta}^{\mbox{\scp open}}(L,\frac{1}{2}) 
\nonumber \\
= L e_{\Delta}^{L,\mbox{\scp open}}=
 Le_{\Delta}^{\infty}(\frac{1}{2})+f_{\Delta}^s - 
\frac{\pi c_{\Delta} v_{\Delta}}{24L}+ o(\frac{1}{L}),
\eea
where $c_{\Delta}=1$, $v_{\Delta}$ is given by \rf{5.11}, and $f_{\Delta}^s$ is 
the surface energy due to the open ends of the lattice. According to conformal 
invariance\cite{ab3q} the leading behavior of the mass gap associated to the 
eigensector with density $n = \frac{1}{2} + \frac{\nu}{L}$ ($\nu = \pm 1, 
\pm 2,\ldots$) of up spins is given by
\bea \label{5.33}
G_L^{\nu,\mbox{\scp open}} = 
E_{\Delta}^{\mbox{\scp open}}(L,\frac{1}{2}+\frac{\nu}{L}) - 
E_{\Delta}^{\mbox{\scp open}}(L,\frac{1}{2}) 
\nonumber \\
= \frac{\pi}{L}v_{\Delta}x_{\Delta}^{s,\nu} + o(\frac{1}{L}),
\eea
where $x_{\Delta}^{s,\nu}$ is a surface critical exponent. The finite-size 
analysis of the model\cite{ab3q} gives the exact result
\beq \label{5.34}
x_{\Delta}^{s,\nu} = 2 x_{\Delta}\nu^2,
\eeq
where $x_{\Delta}$ is given by \rf{4.3}. Comparing \rf{5.33} with 
\rf{5.25} and using \rf{5.34} we verify that the lower mass gaps 
($\nu= \pm1, \pm2, \ldots$) does not depend on the boundary condition, 
up to order $(1/L)$.

We next verify whether the exact results \rf{5.32} and \rf{5.33}-\rf{5.34} 
can be obtained from the LDA functionals introduced in Sec. \ref{secIII}, 
whose infinite-size and finite-size versions now become
\beq \label{5.35}
W_{\Delta}^{\mbox{\scp LDA},L,\mbox{\scp open}} (n_1,\ldots,n_L) \equiv 
\sum_{i=1}^L w_{\Delta}^{L,\mbox{\scp open}}(n_i),
\eeq
and
\beq \label{5.36}
W_{\Delta}^{\mbox{\scp LDA},\infty} (n_1,\ldots,n_L) \equiv 
\sum_{i=1}^L w_{\Delta}^{\infty}(n_i).
\eeq
When the latter approximation is applied to an open system, it misses 
corrections of order $L^0$, which arise from the fact that in a finite-size
open system one nearest-neighbor interaction is missing, as compared to the
finite periodic case. In the thermodynamic limit the resulting error is
negligible, but for finite systems it can be approximately corrected by
replacing the preceding equation by
\beq
W_{\Delta}^{\mbox{\scp LDA},\infty} (n_1,\ldots,n_L)
\approx \frac{L-1}{L} \sum_{i=1}^L w_{\Delta}^{\infty}(n_i),
\label{5.36b}
\eeq
which we take as our definition of $W_{\Delta}^{\mbox{\scp LDA},\infty}$
in this case.

The ground state belongs to the special eigensector $n=1/2$, 
where we have a homogeneous distribution of densities of up spins. 
This means that, as in the periodic case, use the finite-size LDA 
\rf{5.35} yields the exact result
\beq \label{5.37}
E^{\mbox{\scp LDA},L,\mbox{\scp open}} (L,\frac{1}{2}) = 
Le_{\Delta}^{L,\mbox{\scp open}}(\frac{1}{2}),
\eeq
implying exact predictions for the conformal anomaly, sound velocity and 
surface energy. It is interesting to mention that in order to obtain the 
exact result \rf{5.37} it was crucial to use 
$e_{\Delta}^{L,\mbox{\scp open}}(n)$ in the LDA \rf{5.35}. If we used 
$(L-1)e_{\Delta}^{L,\mbox{\scp per}}/L$ instead, we would get an inexact 
result, and wrong predictions for the sound velocity and surface energy.

Let us now consider the results for the ground state energy following from
the LDA \rf{5.36}. In this case we can derive the results from those obtained 
from the LDA \rf{5.5}, appropriate to the periodic case. The external magnetic 
field \rf{4.10} used for the Kohn-Sham auxiliary Hamiltonian \rf{4.9} is the 
XY model with open boundaries in the presence of an uniform magnetic field. 
The ground state energy of the Kohn-Sham Hamiltonian is now given by 
\beq \label{5.38}
E_s(h_s) = Le_{\Delta =0}^{L,\mbox{\scp open}} (\frac{1}{2}) - 
Lh^s\frac{1}{2}.
\eeq
By plugging the LDA \rf{5.36} in \rf{4.13} we obtain
\bea \label{5.39}
E^{\mbox{\scp LDA},\infty}(L,\frac{1}{2}) = 
\nonumber \\
L[e_0^{L,\mbox{\scp open}}(\frac{1}{2}) +
\frac{L-1}{L}(e_{\Delta}^{\infty}(\frac{1}{2}) 
-e_{0}^{\infty}(\frac{1}{2}))],
\eea
where we have used \rf{5.38} and the magnetic field given in \rf{5.2}. The 
asymptotic behavior \rf{5.10} and \rf{5.32} gives us, as $L\rightarrow \infty$, 
\bea \label{5.40}
E^{\mbox{\scp LDA},\infty}(L,\frac{1}{2}) = 
L e_{\Delta}^{\infty}(\frac{1}{2})  
\nonumber \\
+[f_0^s - e_{\Delta}^{\infty}(\frac{1}{2}) + 
e_{0}^{\infty}(\frac{1}{2})] 
-\frac{\pi c_0v_0}{24L} + o(\frac{1}{L}).
\eea
Upon comparing \rf{5.40} with the exact result \rf{5.32}, we see that the 
LDA \rf{5.36} gives wrong results for the surface energy, sound velocity and 
conformal anomaly. 

The sound velocity $v_{\Delta}$ and the surface energy $f_s$ are non universal 
quantities. Consequently is not a surprise, due to our experience with 
the periodic case, that the LDA \rf{5.36} does not give exact 
predictions in the present case. The underlying conformal field theory 
governing the fluctuations of the quantum chain is defined on a half plane 
and has a fixed sound velocity and surface energy.\cite{anomaly,cardy} 
We can also obtain a 
quantum chain with this property by considering the rescaled Hamiltonian
\beq \label{5.41}
\tilde{H}^{\mbox{\scp XXZ}}(\Delta) =
({H}^{\mbox{\scp XXZ}}(\Delta) -Le_{\Delta}^{\infty} - 
f_{\Delta}^s)/v_{\Delta},
\eeq
where $H^{\mbox{\scp XXZ}} (\Delta)$ is the Hamiltonian \rf{4.2} with 
open boundary conditions. For this Hamiltonian, the ground state energy 
$\tilde{e}^{\infty} =0$, the sound velocity 
$\tilde{v}_{\Delta}=1$ and the surface energy $\tilde{f}_{\Delta}^s =0$ 
are constants. The infinite-size LDA \rf{5.36}, applied to \rf{5.41}, 
recovers the correct results.

In order to evaluate the energy gaps, in the case of open boundaries, we 
need to consider the lowest eigenenergy in the sector with average density 
$n=\frac{N}{L} \neq \frac{1}{2}$. In this case the density distribution
is not homogeneous anymore and we must diagonalize the Kohn-Sham Hamiltonian
numerically, and solve self-consistently for the densities. These 
calculations are presented in Sec. \ref{secVB}. Before performing these 
calculations, we introduce, in Sec. \ref{secVA},  approximate analytical 
expressions of the functionals. 

\section{Nonlocal approximations for the XXZ chain}\label{secV}

In the preceding section we applied the finite-size and the infinite-size
LDA functionals to systems in which the only source of inhomogeneity are
boundaries. The finite-size LDA, due to its nonlocal dependence on the system
size, turns out to be superior, but even the more conventional infinite-size
LDA produces correct critical exponents, and, for rescaled Hamiltonians, 
also correct conformal anomalies. However, neither LDA is guaranteed to be
reliable for systems in which the density distributions is inhomogeneous
also in the bulk, as is the case in the presence of site-dependent external 
fields $h_i^{\mbox{\scp ext}}$. We now develop and test nonlocal
approximations for such inhomogeneous systems.

\subsection{Construction of nonlocal functionals}\label{secVA}

In order to calculate the ground state energies produced by inhomogeneous 
density distributions it is necessary to solve selfconsistently the Kohn-Sham 
Hamiltonian \rf{4.9} with $h_i^s$ given by \rf{4.10}-\rf{4.12}. The quality 
of the results will depend on the approximation used for the functional 
$W$, i.e., on the number of terms kept in the expansion \rf{3.16}.

In section \ref{secIV} we considered only the first term in \rf{3.16}, 
producing the functionals \rf{5.4}-\rf{5.5}, \rf{5.21}-\rf{5.22} and 
\rf{5.35}-\rf{5.36}, for periodic, twisted and open boundary conditions,
respectively. These functionals are exact for completely
homogeneous systems, and produce reasonable approximations for weakly 
inhomogeneous situations. To improve the results also in more strongly
inhomogeneous cases, we now additionally consider the second term in 
\rf{3.16}. As we shall see, this extra term is necessary for the correct 
evaluation of the mass gaps of the quantum chain with open boundaries. 

On periodic lattices, the functionals exhibit generalized translation
invariance in the sense that for any integer $m$, $F_a(n_1,\ldots,n_L) 
= F_a(n_1+m,\ldots,n_L+m)$, where $a=I,s$, $n_{j+L}=n_j$ ($j=1,\ldots,L$).
In this case the functions $\alpha_1[n] = \cdots =\alpha_L[n]= \alpha[n]$ 
defined in \rf{3.18}
are site independent and the second term in \rf{3.17} reduces to
\beq \label{3.30}
\sum_{j=1}^L \alpha_j[n_j] (n_j-n_i) = L \alpha[n_i](\frac{N}{L}-n_i).
\eeq
From \rf{3.16}-\rf{3.20} we then have
\bea \label{6.1}
W_{\Delta}^{\mbox{\scp LDA},\infty} (n_1,\ldots,n_L) = 
\\
\sum_{i=1}^L w_{\Delta}^{\infty}(n)|_{n= n_i} +
\sum_{i=1}^L L \alpha[n_i](\frac{N}{L} -n_i),
\eea
where $\sum_{i=1}^L n_i = N$, 
$w_{\Delta}^{\infty}(n) = e_{\Delta}^{\infty}(n)-e_0^{\infty}$, 
$e_{\Delta}^{\infty}(n)$ is the 
ground state energy per site of the infinite homogeneous system 
with total density $n$ of up spins, and 
\beq \label{6.2}
\alpha[n_i] = \lim_{\delta \to 0} \frac{1}{L}
\frac
{W_{\Delta}(\ldots,n_i,n_i+\delta,n_i,\ldots) - 
W_{\Delta}(n_i,\ldots,n_i)}{\delta}.
\eeq
In order to estimate $\alpha[n]$ we approximate 
\beq \label{6.3} 
W_{\Delta}(\ldots,n_i,n_i+\delta,n_i,\ldots) \approx 
(L-1)w_{\Delta}^{\infty}(n_i) 
+ w_{\Delta}^{\infty}(n_i + \gamma\delta/2),
\eeq
where $\gamma \approx 1$. The last term gives the contribution due to 
the inhomogeneity $n_i + \delta$. Since the Hamiltonian is composed 
of two-body interactions we expect that the contribution due to 
the inhomogeneity is given by the energy evaluated around  the average density 
$(n_i+(n_i +\delta))/2$, and hence\cite{gamma} $\gamma \approx 1$.
By plugging 
\rf{6.3} in \rf{6.2} we obtain
\beq \label{6.4}
\alpha[n_i] = \frac{1}{L}\frac{\partial w_{\Delta}^{\infty}(n+
\gamma\delta/2)}{\partial \delta} = 
\frac{\gamma}{2L}\frac{\partial w_{\Delta}^{\infty}(n_i)} 
{\partial n_i},
\eeq
and hence,
\bea \label{6.5}
W_{\Delta}^{\mbox{\scp LADA},\infty} (n_1,\ldots,n_L) = 
\nonumber \\
\sum_{i=1}^L w_{\Delta}^{\infty}(n_i) + 
\frac{\gamma}{2}\sum_{i=1}^L (\frac{N}{L} -n_i)
\frac{\partial w_{\Delta}^{\infty}}{\partial n}(n){\Bigg |}_{n = n_i}.
\eea

In the case of open boundaries, differently from the periodic case, the 
number of links on the lattice is $(L-1)$. Moreover, the second term on the 
expansion in \rf{3.17} is now site dependent. An approximate functional for
this case is obtained by taking the contributions of the boundary sites 
$1$ and $L$ to be half that of the other sites $i=2,\ldots,L-1$. 
Similar arguments as those used to obtain \rf{5.36} for the first term and 
\rf{6.3} for the second term in \rf{3.17}, then give the functional 
\bea \label{6.6}
W_{\Delta}^{\mbox{\scp LDA},\infty} (n_1,\ldots,n_L) = 
\frac{L-1}{L}\left[
\sum_{i=1}^L w_{\Delta}^{\infty}(n_i)  \right. \nonumber \\
\left. +\frac{\gamma}{4}\sum_{i=1}^L (2 -\delta_{i,1}-\delta_{i,L})
(\frac{N}{L} -n_i)\frac{\partial w_{\Delta}^{\infty}}{\partial n} 
(n) {\Bigg |}_{n = n_i}\right].
\eea

Approximations \rf{6.5} and \rf{6.6}, for periodic and open chains
respectively, can be used, up to this point, for any quantum chain. Note
that although we started out from a functional depending on the lattice
gradient $n_j-n_i$, the final expressions depend only on the differences
$N/L-n_i$, i.e., on the deviation of the local density from the average
density of the system. This average makes these functionals highly nonlocal,
in a similar way as described below Eq.~(\ref{3.22}) for the finite-size 
LDA. The present functionals, however, depend on $L$ in a more complex way 
than the finite-size LDA, and additionally depend on $N$. For these
reasons we refer to \rf{6.5} and \rf{6.6} as {\em local and average density 
approximation} (LADA), highlighting thus its simultaneous local and nonlocal
dependence on $n_i$. (In fact, there is a remote conceptual similarity to
the average-density approximation of {\em ab initio} DFT,\cite{dftbook}
which also displays a nonlocal dependence on the density via an integral 
of $n({\bf r})$ over a certain region in space.)

To make these functionals useful in practice, it is necessary to 
evaluate the ground state energy per site $e_{\Delta}^{\infty}(n)$ for fixed 
density $n$. Actually this gives us a general procedure to produce a good 
approximate functional for an arbitrary non integrable quantum chain. 
Numerical diagonalization of small chains, through the Lanczos or the 
density matrix renormalization group (DMRG) methods, can give us an  
estimate for the ground state energy as a function of the density. 
This estimate  is then  used to produce the functional \rf{6.5} and {6.6}.
The advantage of studying a chain whose homogeneous limit is 
exactly integrable is precisely the exact evaluation of these 
quantities. For the XXZ chain, they are obtained by solving the 
coupled integral equations \rf{a.1}-\rf{a.3} of the appendix. We could solve 
these equations numerically for a set of discrete densities and use the 
obtained results to numerically define
$e_{\Delta}^{\infty}(n)$ for arbitrary densities $0<n<1$. 
Alternatively, we can use all known information from the exact solution and 
the conformal invariance of the XXZ model, to produce a good analytical 
parametrization, $e_{\Delta}^{\mbox{\scp par}}(n)$, for 
$e_{\Delta}^{\infty}(n)$. Below we follow the second, analytical, route, 
but test the resulting parametrization by comparing it to numerical data.

The Bethe ansatz solution of the XXZ chain at $\Delta=0$ and 
$\Delta \rightarrow -\infty$ indicates that at those couplings the model 
describes spinless non interacting fermions (up spins). 
The model with $\Delta = 0$ is the standard XY model where the up spins 
have a hard-core interaction of unit size, in lattice space units, that 
forbids double occupancy of up spins on one lattice site. The 
model with $\Delta \rightarrow -\infty$, on the other hand, is related to 
an effective model that forbids the occupation of pairs of up spins at 
distances smaller or equal to unity, in lattice space units.\cite{size-note}
Effectively, apart from  harmless constants, these models describe
spinless particles with hard-core sizes $s$ ($s=1$ for $\Delta =0$ 
and $s=2$ for $\Delta \rightarrow -\infty$). In these cases an analytical 
solution, for arbitrary densities $0 <n<1$, can be derived\cite{size}
\bea \label{6.7}
e_{\Delta}^{\infty}(n) =
\nonumber \\
-\frac{2}{\pi} (1-(s-1)n) \sin(\frac{\pi n}{1-(s-1)n}) -
\frac{\Delta}{2}(1-4n),
\eea
where $s=1$ for $\Delta=0$ and $s=2$ for $\Delta \rightarrow -\infty$. 
On the other hand, for $-1\leq \Delta <1$ the model is gapless and the 
ground state belongs to the sector with $N = L/2$ up spins, or average 
density $n=N/L=1/2$ of up spins. Spin-reversal symmetry gives
\beq \label{6.8}
e_{\Delta}^{\infty}(n) =
e_{\Delta}^{\infty}(1-n), \;\;\;\;\; 0<n<1. 
\eeq
Moreover, for arbitrary values of $-1\leq \Delta<1$
conformal invariance implies for the 
eigensectors with $N=\frac{L}{2} + \nu$ ($|\nu| <<L$) of up spins  
the leading finite-size behavior 
\beq \label{6.9p}
e_{\Delta}^{L,\mbox{\scp per}}(\frac{\frac{L}{2}+\nu}{L}) = 
e_{\Delta}^{L,\mbox{\scp per}}(\frac{1}{2}) 
+ \frac{2\pi v_{\Delta}x_{\Delta}\nu^2}{L^2} + o(\frac{1}{L^2}), 
\eeq
or equivalently
\beq \label{6.9}
e_{\Delta}^{L,\mbox{\scp per}}(n) = 
e_{\Delta}^{L,\mbox{\scp per}}(\frac{1}{2}) 
+ 2\pi v_{\Delta}x_{\Delta}(n-\frac{1}{2})^2 + o(\frac{1}{L^2}), 
\eeq
for $n\approx 1/2$. The parameter $x_{\Delta}$ is given by Eq.~\rf{4.3} 
and gives the critical exponents (see Eq.~\rf{4.1}), and $v_{\Delta}$ is 
the sound velocity given by Eq.~\rf{5.11}. This means that 
\beq \label{6.10}
\frac{d e_{\Delta}^{L,\mbox{\scp per}}(n)}{d n} {\Bigg |}_{n=\frac{1}{2}} = 0, 
\;\;\;\; 
\frac{d^2 e_{\Delta}^{L,\mbox{\scp per}}(n)}{d^2 n}{\Bigg |}_{n =\frac{1}{2}}
= 4\pi v_{\Delta}x_{\Delta}. 
\eeq

Collecting the information in Eqs.~\rf{6.7},\rf{6.8} and \rf{6.10} we obtain 
an approximate analytical parametrization of the ground state energy at 
arbitrary densities for $-1\leq \Delta <1$
\bea \label{6.11}
e_{\Delta}^{\mbox{\scp par}}(n) = 
-\frac{2}{\pi}(1-(s-1)n)\sin(\frac{\pi n}{1 -(s-1)n}) 
\nonumber \\
- \frac{\Delta}{2}
(1-4n) +c_1 +\frac{1}{2}c_2(n-\frac{1}{2})^2, \;\;\; (n<\frac{1}{2}),
\eea
and
\beq \label{6.12}
e_{\Delta}^{\mbox{\scp par}}(n) = e_{\Delta}^{\mbox{\scp par}}(1-n), 
\;\;\;   
 (n>\frac{1}{2}).
\eeq
The parameter $s$ is obtained by imposing the first condition in \rf{6.10}, 
i.e., by solving the equation 
\beq \label{6.13}
\frac{s-1}{\pi}\sin(\frac{\pi}{3-s}) - \frac{2}{3-s}\cos(
\frac{\pi}{3-s}) + \Delta = 0.
\eeq
The parameter $c_2$ is obtained from the second condition in \rf{6.10}, 
namely
\beq \label{6.14}
c_2 = 4\pi v_{\Delta}x_{\Delta} - \frac{16 \pi}{(3-s)^3} 
\sin(\frac{\pi}{3-s}).
\eeq
Finally, the parameter $c_1$ (a harmless constant for the evaluation of gaps) 
can be obtained by imposing that at $n=\frac{1}{2}$, $e_{\Delta}^{\mbox{\scp 
par}} (\frac{1}{2})$ coincides with the exact result 
$e_{\Delta}^{\infty} (\frac{1}{2})$. This is obtained by 
solving the Bethe ansatz equation in the appendix with 
$\Lambda \rightarrow \infty$.

The parametrization \rf{6.11}-\rf{6.12} is then obtained from the solution, 
for each $\Delta$, of \rf{6.13} (giving the parameter $s$) and by solving 
the Bethe ansatz equations \rf{a.1}-\rf{a.3} at the density $n = \frac{1}{2}$. 
This procedure  is certainly much simpler than the brute force numerical
solution, for each density $n$, of the integral equations \rf{a.1}-\rf{a.3}. 
The parameter $s$ in \rf{6.11}-\rf{6.12} plays the role of a generalized 
interaction range, or effective size, of the hard-core interactions among 
the up spins. It varies from $s=1$ to $s=2$ as $\Delta$ goes from $0$ 
to $-\infty$. As $\Delta$ decreases from zero, this parameter is 
smaller than unity, reflecting the decrease of the repulsion 
among the up spins, since now the static term, controlled by $\Delta$, 
is attractive. In table~\ref{table1} we show, for some values of $\Delta$, 
the parameters entering the parametrization \rf{6.11}-\rf{6.12}.

\begin{table}
\caption{\label{table1} Parameters defining the parametrization
$e_{\Delta}^{ \mbox{\scp par}}$ given in \rf{6.11}-\rf{6.12} for
several values of the anisotropy $\Delta = -\cos \gamma$.}
\begin{ruledtabular}
\begin{tabular}{cccccc}
$\gamma$ & $\Delta $ & $s$ & $c_2$ & $c_1 $ \\ 
$\frac{5\pi}{6}$ & 0.866025 & -1.691654 & 0.326211 & -0.013096 \\ 
$\frac{3\pi}{4}$ & 0.707107 & -0.385388 & 0.444007 & -0.013299 \\ 
$\frac{2\pi}{3}$ & 0.5      &  0.295398 & 0.389751 & -0.009221 \\ 
$\frac{\pi}{2}$  & 0        &  1        & 0        & 0         \\ 
$\frac{\pi}{3}$  & -0.5     &  1.349051 & 0.323765 & -0.003210 \\ 
$\frac{\pi}{4}$  & -0.707107&  1.454903 & 1.133043 & -0.010000 \\ 
$\frac{\pi}{6}$  & -0.866025&  1.527244 & 2.395693 & -0.017932  \\
0                & -1       &  1.583761 & 5.625124 & -0.026728 \\
\end{tabular}
\end{ruledtabular}
\end{table}

\begin{figure}[ht!]
\centering
{\includegraphics[angle=0,scale=0.46]{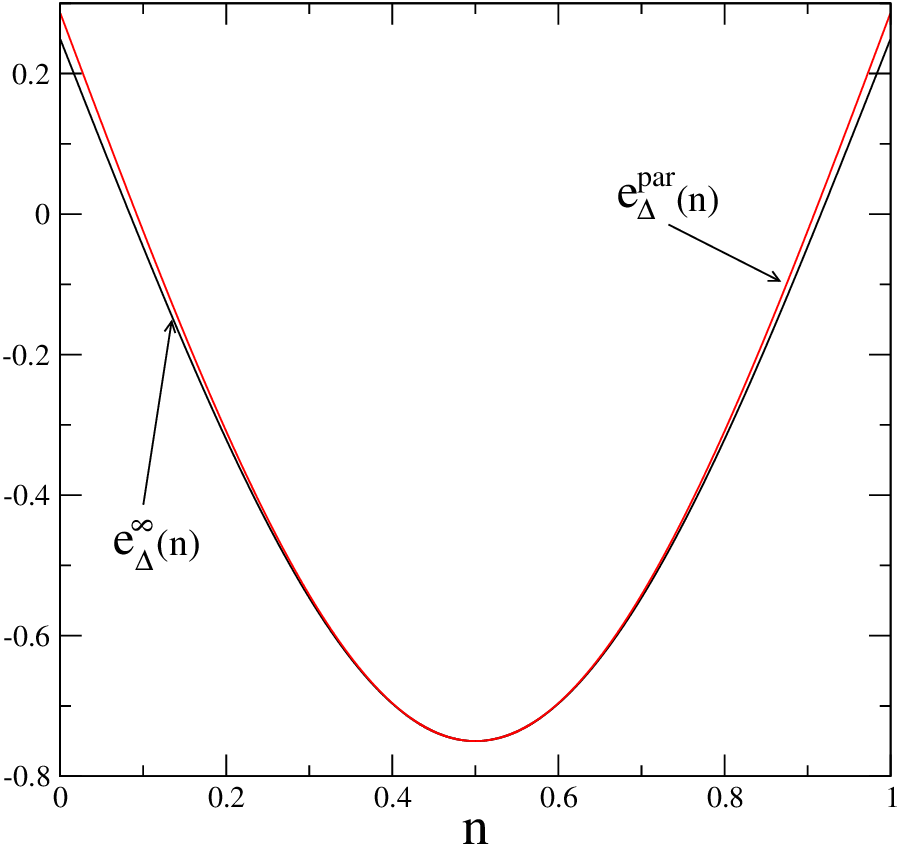}}
\caption{(Color online) 
Parametrization of $e_{\Delta}^{ \mbox {\scp  par}}(n)$, given by
Eqs.~\rf{6.11}-\rf{6.12}, compared to the exact ground state energy,
$e_{\Delta}^{\infty}(n)$, for anisotropy $\Delta = -1/2$.}
\label{fig1}
\end{figure}

\begin{figure}[ht!]
\centering
{\includegraphics[angle=0,scale=0.46]{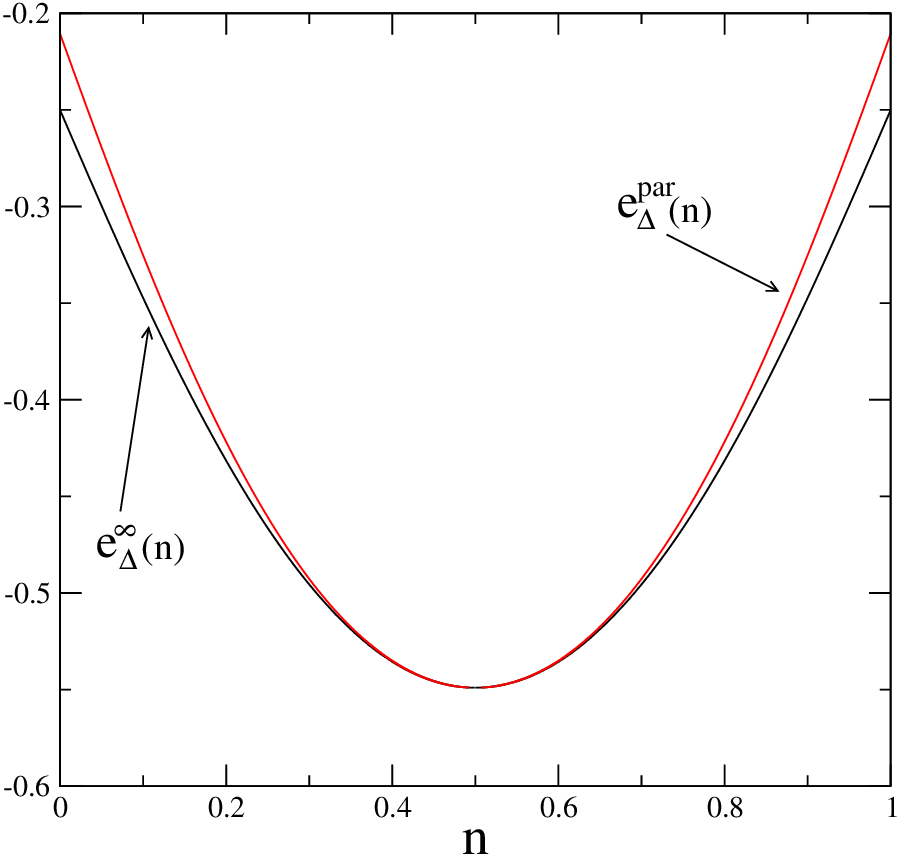}}
\caption{(Color online) 
Parametrization of $e_{\Delta}^{\mbox {\scp par}}(n)$, given by
Eqs.~\rf{6.11}-\rf{6.12}, compared to the exact ground state energy 
$e_{\Delta}^{\infty}(n)$ for anisotropy $\Delta = 1/2$.}
\label{fig2}
\end{figure}

   In order to compare qualitatively our parametrization \rf{6.11}-\rf{6.12} 
with the exact result we show in Fig.~\ref{fig1} and Fig.~\ref{fig2} 
$e_{\Delta}^{\infty}(n)$ and $e_{\Delta}^{\mbox{\scp par}}
(n)$ at anisotropies $\Delta = -\frac{1}{2}$ and $\Delta = \frac{1}{2}$. 
We see from these figures that for $0.4 <n<0.6$ the largest deviation is 
around $0.2\%$. This means that for inhomogeneous density distributions 
$\{n_i,i=1,\ldots,L\}$ with $0.4<n_i<0.6$ the exact result 
$e_{\Delta}^{\infty}(n)$ can be replaced for our parametrization 
$e_{\Delta}^{\mbox{\scp par}}$ of the functionals \rf{6.5} and \rf{6.6}, 
with excellent accuracy. For large inhomogeneities one must numerically solve 
the Bethe ansatz equations, requiring additional computational effort. 

\begin{figure}[ht!]
\centering
{\includegraphics[angle=0,scale=0.46]{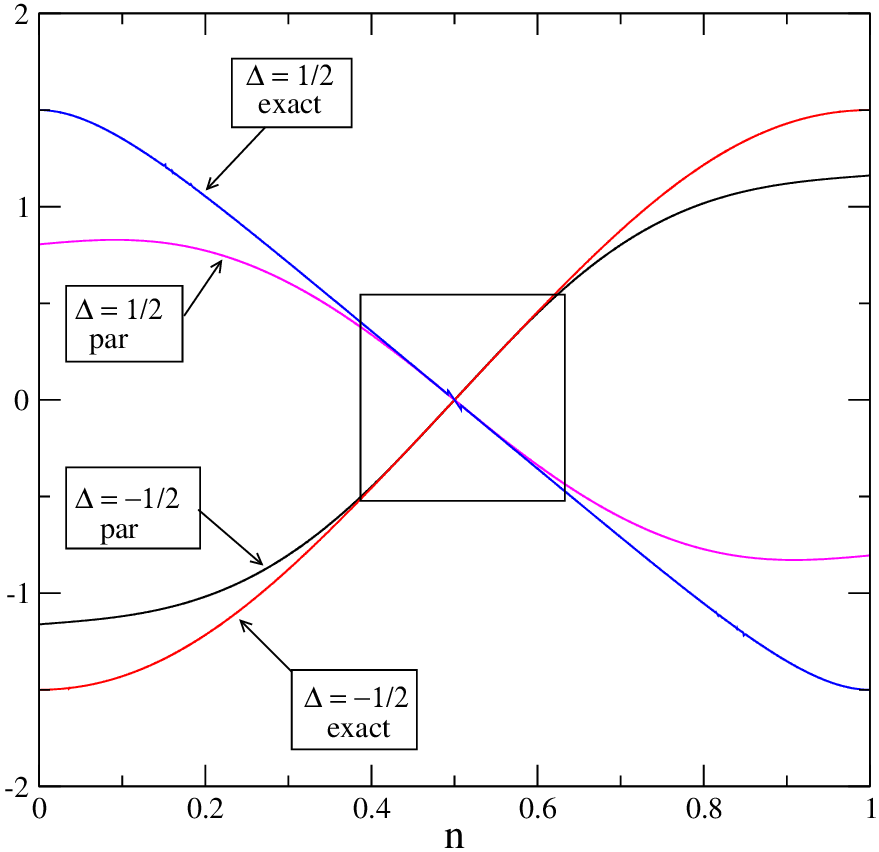}}
\caption{(Color online) 
Effective field $h^s=h_1^s=\cdots=h_L^s$ obtained from \rf{6.15} with
$h_i^{\mbox {\scp ext}} =0$ as a function of the density $n=n_1=\cdots=n_L$ 
for two values of the anisotropies. The exact and approximate curves are 
obtained by using in \rf{6.15} the parametrization \rf{6.11}-\rf{6.12} and 
the exact result, respectively.}
\label{fig3}
\end{figure}

It is important to stress that the density of the interacting system is
reproduced via the site-dependent magnetic fields $\{h_i^s\}$ of the 
non-interacting 
Kohn-Sham Hamiltonian \rf{4.9}. These effective fields depend on the 
difference of derivatives of the functionals at the anisotropy $\Delta$ 
and $\Delta=0$, i.e., 
\beq \label{6.15}
h_i^s = h_i^{\mbox{\scp ext}} + \frac{d}{d n} 
W_{\Delta}^{\mbox{\scp LDA},\infty} (n)|_{n=n_i}.
\eeq
In order to compare the effective magnetic fields produced by our 
parametrization $e_{\Delta}^{\mbox{\scp par}}(n)$ with those obtained  
from the exact values  for $e_{\Delta}^{\infty}(n)$, we show in Fig.~\ref{fig3} these 
fields, for $h_i^{\mbox{\scp ext}} =0$, for two values of the anisotropies,
in a system with constant density $n_1=\cdots= n_L=n$. In this case 
$h_1^s,\ldots,h_L^s=h^s$. 
The exact results are obtained within computer accuracy 
from the numerical solutions of the Bethe 
ansatz equations given in the appendix, and the derivatives were obtained 
by using a cubic spline fitting of the numerical data. We see from this figure 
that for $0.4 < n < 0.6$ our parametrization \rf{6.11}-\rf{6.12} produces a 
rather good approximation for the local potentials in the 
Kohn-Sham Hamiltonian. 

\begin{figure}[ht!]
\centering
{\includegraphics[angle=0,scale=0.46]{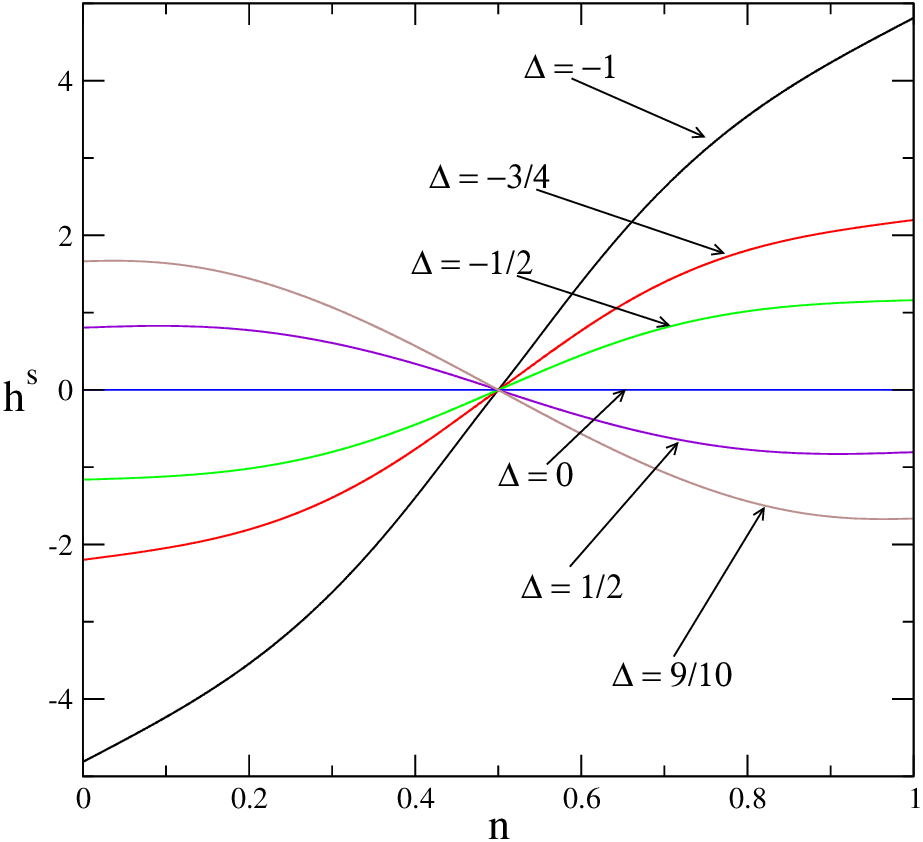}}
\caption{(Color online) 
Effective field $h^s =h_1^s=\cdots=h_L^s$ obtained from \rf{6.15} with 
$h_i^{\mbox {\scp ext}} =0$ as a function of the density $n=n_1=\cdots=n_L$, 
for several values of the anisotropies. The curves were obtained by using the 
parametrization \rf{6.11}-\rf{6.12} in \rf{6.6} and \rf{6.15}.}
\label{fig4}
\end{figure}

It is interesting to observe from Fig.~\ref{fig3} that the magnitude and even 
the sign of $h^s$ depend on the site density and on the value of the anisotropy 
$\Delta$. In Fig.~\ref{fig4} we show the effective magnetic field $h^s(n)$ 
obtained from the parametrization \rf{6.10}-\rf{6.11} for several values of 
the anisotropy $\Delta$. We see from this figure that while for $\Delta <0$ 
the magnetic field $h^s$ increases with the density $n$ of up spins, 
for $\Delta >0$ we have the opposite behavior. This is quite reasonable, since 
for $\Delta <0$ ($\Delta>0$) the net effect of the anisotropy is to increase 
(decrease) the energy as the density increases and the up spins are forced, 
in competition with the kinetic energy, to occupy nearest neighbor sites. 

Next, we apply the LADA functionals \rf{6.5} and \rf{6.6}, with 
\rf{6.10}-\rf{6.11}, to several quantum chains whose ground state 
exhibits a spatially inhomogeneous distribution of up spins. 
Such inhomogeneities can be produced either by boundary conditions 
or by the presence of inhomogeneous external fields. We are going to 
consider example of both cases in the following subsections.

\subsection{Open boundary conditions}\label{secVB}

As discussed in Sec. \ref{secIVC}, except for the eigensector with total 
density $n =1/2$ of up spins ($L$ even), the eigenstates of the XXZ 
chain with open boundary conditions produce inhomogeneous density distributions
even in the absence of external fields.

We have used the LADA approximation \rf{6.6}, with 
$e_{\Delta}^{\infty}(n)$ approximated by the parametrization for
$e_{\Delta}^{\mbox{\scp par}}(n)$ given by \rf{6.11}-\rf{6.14}, to obtain
the effective magnetic fields $\{h_1^s,\ldots,h_L^s\}$ entering the Kohn-Sham 
Hamiltonian \rf{4.9}. 
Through a Jordan-Wigner transformation\cite{jordan-wigner} this Hamiltonian 
is easily diagonalized for lattice sizes up to $L \approx 3000$. The density 
of up spins in the ground state is then calculated 
by diagonalizing \rf{4.9} with \rf{6.15}. In order to compare our 
LADA predictions with exact results for small chains, we present in 
table~\ref{table2}  
some of our spectral calculations at anisotropy $\Delta = -1/2$.
The table displays the lowest energies $E_{\Delta}^{\mbox{\scp open}}(L,
\frac{1}{2} +\frac{\nu}{L})$ and the mass gaps $G_L^{\nu}$ (see \rf{5.33}) 
in the sectors with $N=\frac{L}{2} + \nu$ up spins ($\nu=0,1,2$) and lattice 
sizes $L=4$ to $24$. The exact results were obtained by a direct 
diagonalization of \rf{4.4}, and $\gamma=1$ was used in the LADA functional 
\rf{6.6}. We observe good agreement of the LADA predictions with the exact 
data, becoming better as the lattice size increases. 

\begin{table*}
\caption{\label{table2} Lowest eigenenergies in the sector with
$N= \frac{L}{2} + \nu$ ($\nu =0,1$) up spins of the XXZ chain with open
boundaries with $\Delta =-\frac{1}{2}$ and $L=4-24$. The 2nd and 4th (3rd
and 5th) columns are the exact results (LADA results). The exact and LADA
results for the gaps are shown in the 6th and 7th columns, respectively.}
\begin{ruledtabular}
\begin{tabular}{ccccccl}
L &  $E_{\Delta}^{\mbox{\scp open}}(L,\frac{1}{2})$ &
     $E^{\mbox{\scp LADA},L,\mbox{\scp open}}(L,\frac{1}{2})$ &
     $E_{\Delta}^{\mbox{\scp open}}(L,\frac{1}{2}+\frac{1}{L})$ &
     $E^{\mbox{\scp LADA},L,\mbox{\scp open}}(L,\frac{1}{2}+\frac{1}{L})$ &
    $LG_L^{1,\mbox{\scp open}}$ &
    $LG_L^{1,\mbox{\scp LADA,open}}$ \\
4  & -0.678043  & -0.644052  & -0.437500 & -0.387850 & 0.962172 & 2.624810 \\
6  & -0.699296  & -0.676810  & -0.581496 & -0.549768 & 0.706798 & 0.762252 \\
8  & -0.710812  & -0.694054  & -0.640983 & -0.618987 & 0.558637 & 0.600536 \\
10 & -0.718050  & -0.704710  & -0.671853 & -0.655200 & 0.461971 & 0.495100 \\
12 & -0.723025  & -0.711951  & -0.690197 & -0.676901 & 0.393828 & 0.420600 \\
14 & -0.726656  & -0.717194  & -0.702125 & -0.691102 & 0.343425 & 0.365288 \\
16 & -0.729424  & -0.721166  & -0.710396 & -0.700999 & 0.304446 & 0.321712 \\
18 & -0.731604  & -0.724279  & -0.713276 & -0.708232 & 0.273446 & 0.288846 \\
20 & -0.733365  & -0.726786  & -0.720956 & -0.713716 & 0.248198 & 0.261400 \\
22 & -0.734819  & -0.728847  & -0.724490 & -0.717998 & 0.227235 & 0.238678 \\
24 & -0.736039  & -0.730572  & -0.727307 & -0.721423 & 0.209568 & 0.219576 \\
\end{tabular}
\end{ruledtabular}
\end{table*}

As discussed in Sec.~\ref{secIV}, the mass-gap amplitudes of the finite-size 
corrections give the surface critical exponents $x_{\Delta}^{s,\nu}$ of the 
models (see Eq.~\rf{5.33}). We verified, for arbitrary values of $\Delta$, that 
the LADA functional \rf{6.6} with $\gamma=1$ and the parametrization \rf{6.11} 
gives quite good predictions for the surface exponents. In tables~\ref{table3} 
and \ref{table4} we show some of our estimates obtained for lattice sizes 
up to $L= 512$. The finite-size estimates for the exponent 
$x_{\Delta}^{s,\nu}$ with $\nu=1$ and the ratio 
$x_{\Delta}^{s,2}/x_{\Delta}^{s,1}$ are presented in the last columns of these 
tables. Exact results in the thermodynamical limit $L\rightarrow \infty$ 
are shown in the last line of the tables. These results were derived from 
the Bethe ansatz equations (see appendix) and conformal invariance 
(see \rf{5.34} and \rf{4.3}).

\begin{table*}
\caption{\label{table3} Lowest eigenenergies in the sector with $N=\frac{L}{2} 
+ \nu$ ($\nu =0,1,2$) up spins of the XXZ chain with open
boundaries and anisotropy $\Delta =- \frac{1}{2}$.
The results were obtained by taking $\gamma =1$ and using the
parametrization \rf{6.11}-\rf{6.12} in the LADA \rf{6.6}. In the
5th and 6th columns we show the finite-size estimates for
the surface critical exponents $x_{\Delta}^{s,1}$ and the
ratios $x_{\Delta}^{s,2}/x_{\Delta}^{s,1}$, respectively. The last line gives
the expected exact results in the bulk limit ($L \rightarrow \infty$).}
\begin{ruledtabular}
\begin{tabular}{cccccc}
 L               &  $a=E^{\mbox{\scp open}}_{\Delta}(L,\frac{1}{2})/L$ &
$b=E^{\mbox{\scp open}}_{\Delta}(L,\frac{1}{2}+\frac{1}{L})/L$ &
$c=E^{\mbox{\scp open}}_{\Delta}(L,\frac{1}{2}+\frac{2}{L})/L$
& $(b-c)L^2/\pi$ &
 $\frac{c-a}{b-a}$ \\
16 & -0.721166 & -0.700999 & -0.641516 & 1.643310 & 3.949620
\\
32 &-0.735349 & -0.730155 & -0.714603 & 1.693019 & 3.994144
\\
64 & -0.742614 & -0.741299 & -0.737349 & 1.714280 & 4.003604
\\
128 & -0.746291 & -0.745961 & -0.744968 & 1.723679 &
 4.003837
\\
256 &-0.748142 & -0.748059 & -0.747810 & 1.728007 & 4.002522
\\
512 &-0.749070 & -0.749049 & -0.748987 & 1.730070 &4.001427
\\
1024 &-0.749535 & -0.749529 & -0.749513 & 1.731081 &4.000765
\\
1224 &-0.749611 & -0.749607 & -0.749596 & 1.731302 &4.000506
\\
1424 &-0.749665 & -0.749663 & -0.749655 & 1.731408 &4.000439
\\
${\infty \atop {({\rm exact})}}$ & -0.75 & -0.75 & -0.75 & 1.732051 & 4 \\
\end{tabular}
\end{ruledtabular}
\end{table*}

It is interesting, at this point, to discuss the effect of the parameter 
$\gamma$ in our LADA functional \rf{6.6} (or \rf{6.5}). Strictly speaking, 
$\gamma$ is a phenomenological parameter whose value was argued to be 
$\gamma \approx 1$ in Sec. \ref{secVA}. 
The results presented in table~\ref{table3} 
and \ref{table4} were obtained by setting $\gamma =1$. We have also calculated 
the quantities presented in table~\ref{table3} and \ref{table4} for
other values of $\gamma$. For $\Delta <0$ we got good results even from
$\gamma =0$, where the nonlocal correction vanishes and the LADA functional
reduces to the LDA. However for $\Delta >0$ (ferromagnetic regime) we get 
non-vanishing gaps for $\gamma=0$. This means that in this regime nonlocality
is important for inhomogeneous density profiles and cannot be neglected. 
Good results are obtained for all values of $-1 \leq \Delta <1$, as long 
as $0.98 <\gamma< 1.03$, which is consistent with our {\em a priori}
expectation $\gamma \approx 1$.

\begin{table*}
\caption{\label{table4} Same as table \ref{table3}, but for anisotropy
value $\Delta = +\frac{1}{2}$.}
\begin{ruledtabular}
\begin{tabular}{cccccc} 
 L               &  $a=E^{\mbox{\scp open}}_{\Delta}(L,\frac{1}{2})/L$ &
$b=E^{\mbox{\scp open}}_{\Delta}(L,\frac{1}{2}+\frac{1}{L})/L$ & 
$c=E^{\mbox{\scp open}}_{\Delta}(L,\frac{1}{2}+\frac{2}{L})/L
$
& $(b-c)L^2/\pi$ &
 $\frac{c-a}{b-a}$ \\
16 & -0.532764 & -0.528385 & -0.513800 & 0.35684 & 4.330570
\\
32 &-0.540667 & -0.539536 & -0.535933 & 0.368617 & 4.186629
\\
64 &-0.544792 & -0.544495 & -0.543580 & 0.386752 & 4.084184
\\
128 &-0.546900 & -0.546822 & -0.546587 & 0.403868 & 4.029560
\\
256 &-0.547865 & -0.547945 & -0.547885 & 0.416341 & 4.008144
\\
512 &-0.548500 & -0.548495 & -0.548480 & 0.424043 & 4.001793
\\
1024 &-0.548769 & -0.548768 & -0.548764 & 0.428364 & 4.000165
\\
1224 &-0.548813 & -0.548812 & -0.548809 & 0.429098 & 4.000075
\\
1424 &-0.548845 & -0.548844 & -0.548842 & 0.429645 & 3.999887
\\
${\infty \atop {({\rm exact})}}$ & -0.549038 & -0.549038 & -0.549038 & 0.4330131 & 4 \\
\end{tabular}
\end{ruledtabular}
\end{table*}

\begin{figure}[ht!]
\centering
{\includegraphics[angle=0,scale=0.46]{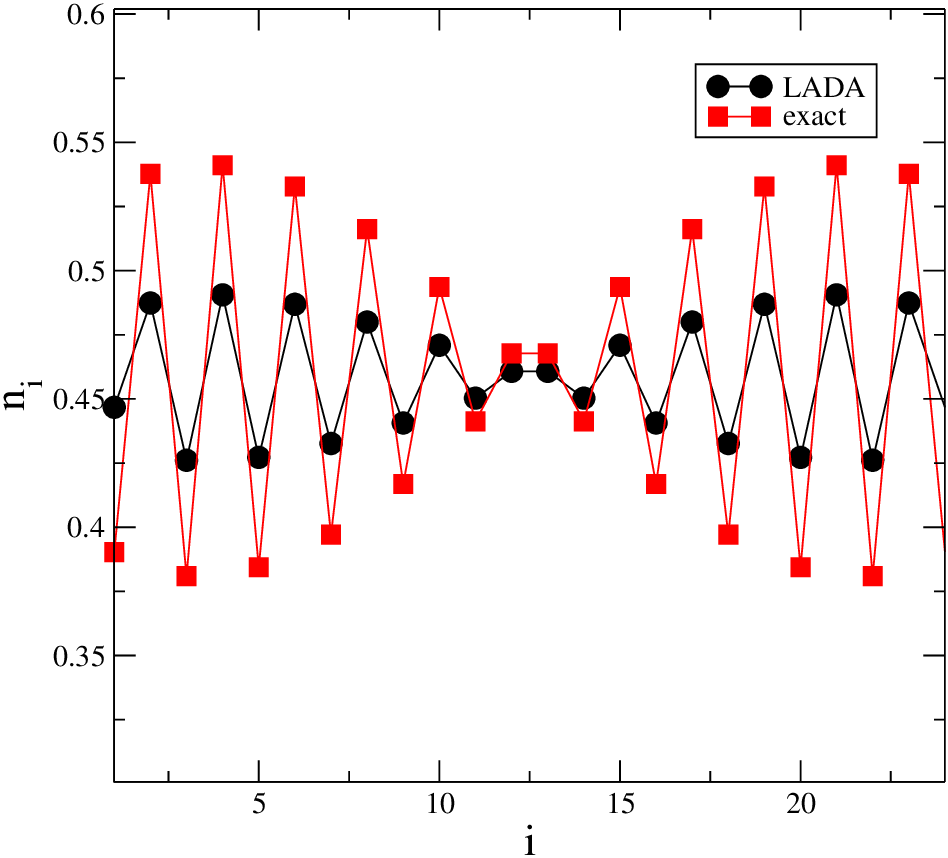}}
\caption{(Color online) 
Exact and LADA predictions for the density profile $n(i)$ of up spins for 
the XXZ quantum chain with open boundaries and $\Delta = -\frac{1}{2}$. 
The profiles correspond to the lowest eigenenergy in the sector with 
$N=11$ up spins of the quantum chain with $L=24$ sites. }
\label{fig5}
\end{figure}

\begin{figure}[ht!]
\centering
{\includegraphics[angle=0,scale=0.46]{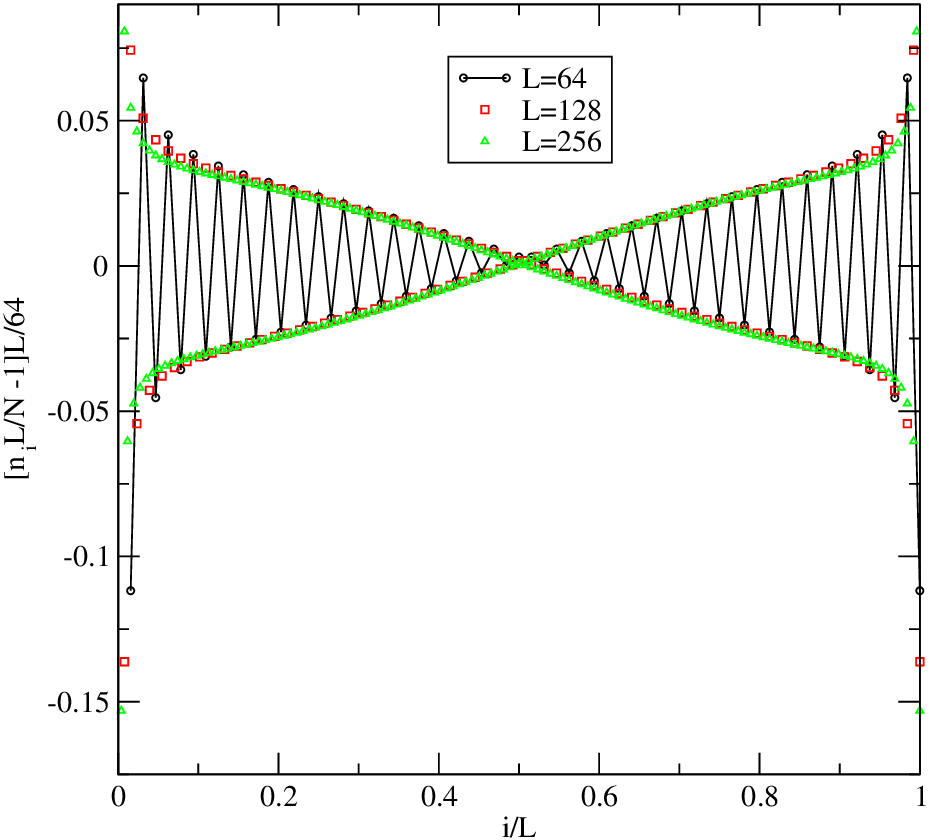}}
\caption{(Color online) 
LADA predictions for the density profiles $n(i)$. The profiles 
correspond to the lowest eigenenergy in the sector with $N= \frac{L}{2}-1$ 
up spins of the XXZ quantum chain with $L$ sites ($L=64,128,256$), 
open boundaries and anisotropy $\Delta = 
-\frac{1}{2}$.}
\label{fig6}
\end{figure}

Let us now consider the density profiles $\{n_1,\ldots,n_L\}$ of up spins on 
finite lattices. The Bethe ansatz for the finite chains, which would give 
the exact results for the density profiles, is impractical to solve. We have 
therefore considered lattice sizes we are able 
to diagonalize numerically. In Fig~\ref{fig5} we show, for $L=24$ sites, an 
example of the exact and LADA prediction for the density profile 
$\{n_1,\ldots,n_{24}\}$ of the XXZ chain with open boundaries. The densities 
shown in the figure are obtained from the lowest eigenenergy in the sector 
with $n=11$ up spins and anisotropy $\Delta = -\frac{1}{2}$. We see from this 
figure that although the LADA amplitudes are smaller than the exact ones,
they exhibit the same spacial oscillations. These spacial oscillations should 
decrease in amplitude as the lattice size increases. In Fig~\ref{fig6} we show 
the density profiles predicted by our LADA functional \rf{6.6} with the 
parametrization function \rf{6.11} and $\gamma =1$, for several lattice sizes. 
The density profiles correspond to the lowest eigenstate in the sector with 
$N= \frac{L}{2}+1$ spins at anisotropy $\Delta = -\frac{1}{2}$. We have scaled 
the vertical and horizontal axis of Fig.~\ref{fig6}, in order to display
simultaneously the density profiles of distinct lattice sizes. This
figure shows that the amplitude of the oscillations decreases as $O(L^{-1})$, 
as the lattice size increases. 

\subsection{Periodic boundary condition with inhomogeneous external magnetic 
fields}\label{secVC}

A physically distinct way to produce inhomogeneous density profiles 
$\{n_1,\ldots,n_L\}$ of up spins is the presence of external magnetic fields 
$\{h_1^{\mbox{\scp ext}},\ldots,h_L^{\mbox{\scp ext}}\}$. In this case, as in 
the case of open boundaries, we have to solve the Kohn-Sham Hamiltonian 
\rf{4.9}-\rf{4.10} selfconsistently.

\begin{figure}[ht!]
\centering
{\includegraphics[angle=0,scale=0.46]{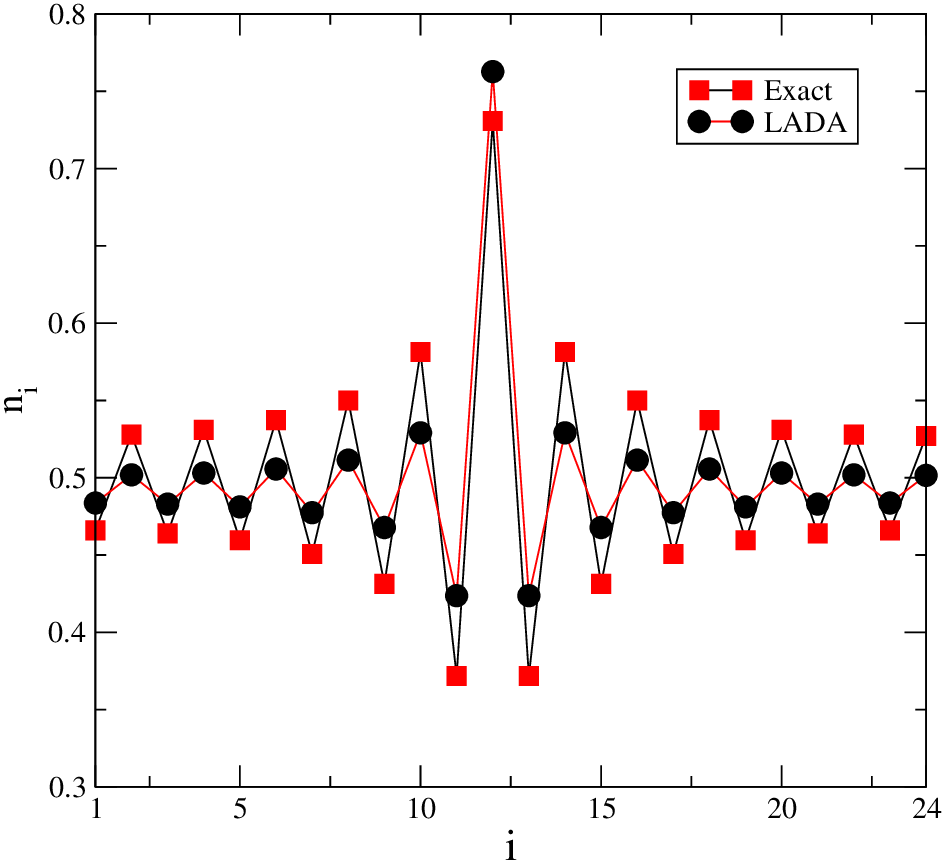}}
\caption{(Color online) 
Exact and LADA results for the density profiles corresponding to the 
ground state of the XXZ quantum chain with periodic boundaries and 
inhomogeneous magnetic fields. The profiles correspond to the quantum chain 
with $L=24$ sites, $N=12$ up spins, anisotropy $\Delta = -\frac{1}{2}$, and 
inhomogeneous magnetic fields $h_i^{\mbox {\scp ext}} = \delta_{i,11}$ 
($i=1,\ldots,24$).}
\label{fig7}
\end{figure}

In order to compare our results to the exact ones, we show in Fig.~\ref{fig7} 
the density profile obtained by using the LADA functional \rf{6.5} 
with $e_{\Delta}^{\infty}(n)$ replaced by 
$e_{\Delta}^{\mbox{\scp par}}(n)$ given by \rf{6.11}. 
The density profiles of Fig.~\ref{fig7} correspond to the lowest energy 
in the sector with $N=12$ up spins  of the periodic XXZ chain with 
$L=24$ sites and anisotropy $\Delta = -\frac{1}{2}$. The external local 
magnetic fields are chosen to be $h_i^{\mbox{\scp ext}} =\delta_{i,11}$. 
Similar to Fig.~\ref{fig5}, the LADA densities have smaller amplitudes than 
the exact ones, but exhibit the same spacial oscillations. 

One of the main purposes of this paper is to present density functionals 
that are able to predict the critical exponents of critical chains from
finite-size corrections to the mass gap. In order to test the LADA
functional \rf{6.5} with \rf{6.11} we study the periodic XXZ chain with a 
single impurity, represented by the external magnetic field 
$h_i^{\mbox{\scp ext}} = \delta_{i,1}$ ($i=1,\ldots,L$). The presence of 
such impurity breaks spin reversal symmetry, but we expect, as 
$L\rightarrow \infty$, the leading behavior for the average mass gap
\bea \label{7.1}
\bar{G}_L^{\nu\mbox{\scp per}} = \frac{1}{2}[
E_{\Delta}^{\mbox{\scp per}}(L,\frac{1}{2} + \frac{\nu}{L}) 
+E_{\Delta}^{\mbox{\scp per}}(L,\frac{1}{2} - \frac{\nu}{L}) 
\nonumber \\
-2E_{\Delta}^{\mbox{\scp per}}(L,\frac{1}{2})] 
\\
=\frac{\pi}{L} v_{\Delta}x_{\Delta}^{s,\nu} + o(\frac{1}{L}), 
\eea
where $v_{\Delta} $ is the sound velocity \rf{5.11} and $x_{\Delta}^{s,\nu}$ 
the surface exponents ($\nu=1,2,\ldots$). 
In Eq.~\rf{7.1} $E_{\Delta}^{\mbox{per}}(L,n)$ is the lowest energy in the 
sector with $N=nL$ up spins of the periodic chain with anisotropy 
$\Delta$ and $L$ sites. In table \ref{table5} we present the finite-size 
estimates for the critical exponents $x_{\Delta}^{s,\nu}$ of the 
quantum chain with $\Delta = - \frac{1}{2}$ and lattice sizes up to $L = 
1024$. 
In the last line we show the expected exact results in the bulk limit
$L \rightarrow \infty$. Clearly, the LADA functional \rf{6.5} with 
parametrization \rf{6.11} gives reliable results for the critical exponents.

\begin{table}
\caption{\label{table5} LADA predictions for the ground state energy
and finite-size estimates of the average mass gaps \rf{7.1} for
the XXZ chain with periodic boundaries and inhomogeneous magnetic fields.
The results were derived for the anisotropy value $\Delta = -\frac{1}{2}$ and
inhomogeneous fields $h_i^{\mbox {\scp ext}} = \delta_{i,L/2}$ ($i=1,
\ldots,L$). In the last line it is shown the expected results in the bulk
limit ($L \rightarrow \infty$).}
\begin{ruledtabular}
\begin{tabular}{ccccc}
 L               &
 $E_{\Delta}^{\mbox{\scp per}}(L,\frac{1}{2})$ &
$L{\bar G}_L^{1,\mbox{\scp per}}\pi$  &
$L{\bar G}_L^{2,\mbox{\scp per}}\pi$  &
${\bar G}_L^{2,\mbox{\scp per}}/{\bar G}_L^{1,\mbox{\scp per}}$ \\
    8 & -0.736502 & 1.646049 & 6.211188 & 3.773392 \\
   16 & -0.738800 & 1.685301 & 6.657321 & 3.950226   \\
   32 & -0.743287 & 1.707394 & 6.816070 & 3.992090   \\
   64 & -0.746363 & 1.719601 & 6.879476 & 4.000623   \\
  128 & -0.748111 & 1.725632 & 6.905708 & 4.001843   \\
  256 & -0.749038 & 1.728826 & 6.917514 & 4.001278   \\
  512 & -0.749515 & 1.730444 & 6.923041 & 4.000730   \\
 1024 & -0.749756 & 1.731253 & 6.925686 & 4.000389   \\
${\infty \atop {({\rm exact})}}$ & -0.75 & 1.732051 & 6.928204 & 4 \\
\end{tabular}
\end{ruledtabular}
\end{table}

\section{Conclusions}\label{secVI}

In this work we explored the possibility to obtain reliable results for 
critical exponents, conformal anomalies, and related quantities of quantum 
chains from density functionals. Due to conformal invariance, the critical 
exponents are obtained from the $O(1/L)$ finite-size corrections of the mass 
gaps of the quantum chains in finite geometries. These finite-size
corrections can be obtained approximately from DFT.
In order to test our general approach, we performed an extensive study of 
various LDA and beyond LDA functionals for the exactly integrable XXZ quantum 
chain in the critical regime $-1 \leq \Delta <1$. 

Our functionals are obtained from a formal gradient expansion of the unknown 
exact functional around homogeneous distributions of the infinite system. 
The first term of this expansion gives LDA functionals, appropriate for cases 
where the density distribution in the chain is weakly inhomogeneous, or fully
homogeneous. The latter is the case when the quantum chain is defined on 
translationally invariant lattices, with periodic and twisted boundary 
conditions (see Eqs.~\rf{5.5} and \rf{5.22}). As we saw in Sec. \ref{secIV}, 
even a simple LDA can give exact critical exponents, but a direct evaluation 
of the conformal anomaly from these functionals gives wrong results because
the sound velocity, a nonuniversal constant, has distinct values for the 
interacting Hamiltonian and the non-interacting auxiliary Kohn-Sham 
Hamiltonian. A correct LDA prediction of the conformal 
anomaly $c=1$, for any value of the anisotropy $\Delta$, can be obtained only 
if we rescale the interacting and non interacting quantum chain by its 
appropriate sound velocity. This is certainly a problem for general quantum 
chains whose sound velocity is unknown. A possible solution is the use of a 
size-dependent LDA, which embodies nonlocal corrections in a simple way.

In the case of open boundaries, only eigenstates belonging to the sector 
with $N= L/2$ up spins have a homogeneous distribution on the 
quantum chain with finite size $L$.
In this case, the surface energy of the quantum chain is predicted 
incorrectly, for the same reason discussed above for the
sound velocity and conformal anomaly. However, as we showed in Sec. 
\ref{secIV} (see Eq.~\rf{5.41}), application to a properly rescaled Hamiltonian 
gives correct results, as does use of a finite-size LDA.

If  the density distribution is inhomogeneous, our results indicate that we 
must also consider at least the second term of the expansion 
\rf{3.16}-\rf{3.18} of the exact functional. 
The LADA functional derived from this term for 
quantum chains with periodic and open boundary conditions is given by \rf{6.5} 
and \rf{6.6}, respectively. These functionals can be used, in principle, for a 
general quantum chain, by replacing $e_{\Delta}^{\infty}(n)$ 
by an approximate expression $e^{\mbox{\scp hom}}(n)$ for the ground state 
energy per site of the infinite system with density $n$. 
In this general case, the quality of the resulting prediction of critical 
exponents, via evaluation of gaps, depends on the quality of approximations
for $e^{\mbox{\scp hom}}(n)$.  For an exactly integrable chain, like the XXZ 
quantum chain, the energy per site $e_{\Delta}^{\infty}(n)$ can be obtained 
exactly, e.g., by solving the integral equation derived from the Bethe ansatz 
(see \rf{a.1}-\rf{a.3} in the appendix).
Instead of using such numerical results for $e_{\Delta}^{\infty}
(n)$ we proposed, in Sec. \ref{secV}, an approximate parametrization
$e_{\Delta}^{\mbox{\scp par}}(n)$, Eq.~\rf{6.11}, containing all 
essential ingredients to furnish good estimates for the critical exponents of 
the XXZ quantum chain in the presence of small inhomogeneities.

Both the finite-size LDA and the LADA functional are nonlocal density 
functionals, the former depending on the local density and the size of the
system, the latter depending on the local and the average density. Unlike
the usual (infinite-size) LDA, these functionals also depend, in a known
way, on the boundary conditions (open, periodic, or twisted). As a consequence
of this nonlocality, the finite-size LDA and the LADA functional each cure
certain defects of the LDA. For finite homogeneous chains, the LDA reproduces 
the correct critical exponents, but predicts a wrong conformal anomaly, 
whereas the finite-size LDA correctly reproduces both. For inhomogeneous 
chains, the LADA functional yields better energies and gaps than the LDA.
For finite homogeneous chains, the LADA functional reduces to the LDA. 
The combination of both improvements, i.e., the construction of a finite-size
LADA is a promissing project for the future.

Another interesting question concerns possible extensions of the LADA 
functional \rf{6.5}-\rf{6.6} to other quantum chains in the $c=1$ universality 
class. Consider a general homogeneous model on 
this universality class. The introduction of a homogeneous magnetic field (or 
chemical potential) $h(n)$ that couples to the density operator $\hat{n}_i$ 
will produce a ground state with energy per site $e(n)$ and homogeneous 
density $n$. In order to fix the density $n$, this magnetic field 
should satisfy $ h(n) = \frac{d e(n)}{d n}$. The critical exponents with 
such external field are given by Eq.~\rf{4.1} with $x=x(n)$, depending 
on the particular density fixed by $h(n)$. Besides $x(n)$ and $h(n)$, the 
quantum chain is also characterized by the non-universal sound velocity 
$v(n)$ that fixes the scale of the momentum-energy dispersion relation. 
The results coming from conformal invariance for this class of models give us 
a generalization of Eq.~\rf{5.14} for arbitrary densities, i.e., 
$\frac{d^2e(n)}{d n^2} = 4\pi v(n)x(n)$. 

In order to obtain the density distribution $\{n_i\}$ when inhomogeneous 
external fields or boundaries are added to the homogeneous system we may 
proceed in two ways. If the local density never exceeds unity, the XY model 
\rf{4.9} can be used as the auxiliary non-interacting (Kohn-Sham) system. 
If this is not the case, the only remaining possibility is the direct use 
of the approximate functional for $F_I$ in the Euler equation \rf{2.7}.
The first case is preferable, since the LDA functional is not exact and the 
auxiliary non-interacting chain will give a more precise evaluation of the 
kinetic energy contained in $F_s$. In both approaches, the necessary 
quantities for the evaluation of the density are $h(n)$ and $4\pi v(n)x(n)$.

To conclude, let us briefly discuss some possible future developments of the 
present work. The functional we produce gives us mass gaps of finite chains 
that produce reasonable, in some cases even exact, estimates for the critical 
exponents. A second possibility to obtain the critical exponents is from the 
power-law decay of the Friedel density oscillations due to the 
presence of impurities on the quantum chain.\cite{friedel} 
Tests with the LADA functionals \rf{6.5} or \rf{6.6} for the XXZ chain 
with open and periodic boundary condition indicate poor results in this 
case. This means that the correct decay of the Friedel 
density oscillations can only be obtained by considering higher order terms 
in the expansion \rf{3.16}-\rf{3.18}. This is certainly an interesting point 
for the future. 

\begin{acknowledgments}
We thank V. L. L\'\i bero for useful discussions. This work was
partially supported by the Brazilian agencies FAPESP and CNPq. 
\end{acknowledgments}

\appendix*
\section{Integral equations derived from the Bethe ansatz solution of the XXZ 
quantum chain}\label{appA}

The ground state energy per site,  $e_{\Delta}^{\infty}(n)$,  
of the XXZ chain, in the bulk limit, can be obtained from the Bethe ansatz 
solution of the model (see e.g. Ref.~\onlinecite{hamer}).

For $-1 \leq \Delta <1$, by setting $\Delta = -\cos \gamma$
\beq \label{a.1}
e_{\Delta}^{\infty}(n) = \frac{\cos \gamma}{2} - 
2\int_{-\Lambda}^{\Lambda} \frac{\sin^2 \gamma}{\cosh \eta - \cos \gamma} 
R(\eta) d\eta,
\eeq
where the parameter $\Lambda$ fix the density $n$ through 
\beq \label{a.2}
\int_{-\Lambda}^{\Lambda}R(\eta)d\eta = n.
\eeq
The function $R(\eta)$ is obtained by solving the integral equation 
\bea \label{a.3}
R(\eta) = 
\frac{1}{2\pi}\left[\frac{\sin \gamma}{\cosh \eta -\cos \gamma} 
\right.  \nonumber \\ \left.
- \int_{-\Lambda}^{\Lambda}\frac{\sin 2\gamma}{\cosh (\eta -\eta') - 
\cos 2\gamma} R(\eta')d\eta'\right].
\eea

\begin{thebibliography}{99}
\bibitem{gaussian-note}In the case of the Gaussian model the Hamiltonian (Euclidean Lagrangean) is given by
$H=-\frac{1}{2}K\sum(\phi -\phi')^2$, where the sum is over all
nearest-neighbor pairs and $\-\infty <\phi <\infty$. In this case
$x=1/4\pi K$.
\bibitem{gamma}In Sec. \ref{secV} we evaluate surface critical exponents using 
various of $\gamma$. Best results are obtained for $\gamma \approx 1$.
\bibitem{size-note}This 
related model has a static term $-\Delta(\sigma_i^z+1)(\sigma_{i+1}^z + 1)/4$ 
that diverges, at $\Delta \rightarrow -\infty$, when two up spins are 
located at nearest neighbor sites.\cite{size}
\end{thebibliography}
\end{document}